\documentclass[11pt,a4paper]{article}
\usepackage{jheppub_kim}

\usepackage{subfigure,amsmath,amssymb ,amsfonts, latexsym}

\usepackage[notcite,color,final
                               ]{showkeys}
\definecolor{refkey}{gray}{.25}
\definecolor{labelkey}{gray}{.25}

       \def\e  {\epsilon}
\def\s  {\sigma}       

\renewcommand{\a}{\alpha}      \renewcommand{\b}{\beta}
\renewcommand{\d}{\delta}      \renewcommand{\k}{\kappa}
\renewcommand{\l}{\lambda}     
      
\renewcommand{\th}{\theta}

\newcommand{\g}{\gamma}      \newcommand{\Th}{\Theta}
 
\newcommand{\m}{\mu}         
\newcommand{\n}{\nu}

\newcommand{\cala}{\mbox{${\cal A}$}} \newcommand{\calb}{\mbox{${\cal B}$}}
 
\newcommand{\cale}{\mbox{${\cal E}$}} \newcommand{\calf}{\mbox{${\cal F}$}}
\newcommand{\calg}{\mbox{${\cal G}$}} 
 \newcommand{\calj}{\mbox{${\cal J}$}}
 \newcommand{\call}{\mbox{${\cal L}$}}
 \newcommand{\caln}{\mbox{${\cal N}$}}
 
 \newcommand{\calr}{\mbox{${\cal R}$}}

\newcommand{\be}{\begin{equation}}
\newcommand{\ee}{\end{equation}}
\newcommand{\beqa}{\begin{subequations}\begin{eqnarray}}
\newcommand{\eeqa}{\end{eqnarray}\end{subequations}}
\newcommand{\nn}{\nonumber}
\newcommand{\LA}[1]{\label{#1}}

\newcommand{\half}{\frac{1}{2}}

\newcommand{\ra}{\rightarrow}

\newcommand{\lra}{\Longrightarrow}

\newcommand{\ud}{\,\mathrm{d}}

\newcommand{\tr}{{\rm tr}\,}

\newcommand{\dd}{\mathrm{d}}
\newcommand{\tE}{{\tilde E}}
\newcommand{\tA}{{\tilde A}}
\renewcommand{\th}{{\tilde h}}
\newcommand{\tB}{{\tilde B}}

\newcommand{\hJ}{{\hat J}}
\newcommand{\tF}{{\tilde F}}

\newcommand{\fg}{\mathfrak{g}}

\newcommand{\nnn}{(I)}

\title{Holographic DC conductivities from the open string metric}

\author[a]{Keun-Young Kim}
\author[b]{and Da-Wei Pang}

\affiliation[a]{ School of Physics and Astronomy, University of
Southampton, \\ Southampton, SO17 1BJ, UK
}
\affiliation[b]{Centro Multidisciplinar de Astrof\'{\i}sica-CENTRA, Departamento de F\'{\i}sica,, \\
 Instituto Superior T\'{e}cnico-IST, Universidade T\'{e}cnica de Lisboa \\
 Av. Rovisco Pais 1, 1049-001 Lisboa, Portugal
}

\emailAdd{k.kim@southampton.ac.uk}
\emailAdd{dawei.pang@ist.utl.pt}


\subheader{SHEP-11-21}

\abstract{
We study the DC conductivities of various holographic models
using the open string metric (OSM), which is an effective metric
geometrizing density and electromagnetic field effect.
We propose a new way to compute
the nonlinear conductivity using OSM.
As far as the final conductivity formula is concerned,
it is equivalent to the Karch-O'Bannon's real-action method.
However, it yields a geometrical insight and technical simplifications. Especially, a real-action condition is interpreted as a regular geometry condition of OSM.
As applications of the OSM method, we study several holographic models
on the quantum Hall effect and strange metal.
By comparing a Lifshitz background and the Light-Cone AdS, we show
how an extra parameter can change the temperature scaling behavior of
conductivity. Finally we discuss how OSM can be used
to study other transport coefficients, such as diffusion constant, and effective temperature induced by the effective world volume horizon.

}

\begin{document}

\maketitle

\section{Introduction}

The AdS/CFT correspondence
has been a powerful tool for studying dynamics
of strongly coupled field theory, as it claims that a weakly coupled classical gravity theory
in certain bulk spacetime is equivalent to a strongly coupled field theory living
on the boundary of that spacetime. It opened up a new window to
understand physics in the real world and has been widely witnessed its tremendous
success in recent years, such as holographic QCD~\cite{Erdmenger:2007cm, CasalderreySolana:2011us}. More recently,
investigations on applications of AdS/CFT to condensed matter theory (often named as AdS/CMT)
have accelerated enormously. See \cite{Hartnoll:2009sz,McGreevy:2009xe,Hartnoll:2011fn}, for example.

One of the important applications of AdS/CFT is to compute transport coefficients in the strongly coupled field theory, such as the shear (bulk) viscosity, diffusion constant and various conductivities.
Of these, in this paper, we focus on the DC conductivity.
The conductivity tensor $\s^{ij}$ measures the response of a conducting medium to an external electric field ($E_j$).
\begin{equation}
  j^i = \s^{ij} E_j
\end{equation}
where $j^i$ are the currents induced in the medium. Anisotropy of medium
or an external magnetic field $B$ can produce an off-diagonal conductivity.
In the context of condensed matter theory this conductivity is particularly interesting in the strange metal phases of heavy fermion compounds and high temperature superconductors.
For example, the strange metal properties are characterized by \cite{Hartnoll:2009ns}
\begin{equation} \label{eq.1}
  \s^{xx} \sim \frac{1}{T} \,, \quad
  \s^{xy} \sim \frac{1}{T^3} \,, \quad
  \cot \theta_H \sim \frac{\s^{xx}}{\s^{xy}} \sim T^2 \,,
\end{equation}
where $\theta_H$ is called the Hall angle, since it is related to the Hall current due to the magnetic field.
The other property is AC conductivity showing $\sigma(\omega)\sim\omega^{-0.65}$,
which we will not pursuit in this paper.  Quantum Hall conductivity also can be addressed by
many holographic models. See, for example, \cite{Bergman:2010gm,Jokela:2011eb} and references therein.
These kinds of DC currents phenomena are governed by strongly-coupled dynamics which
is the the very place that the holographic approach may play a useful role.
After developing a general method, we will apply them to the models
addressing strange metallic conductivity and quantum Hall conductivity.

We want to study the DC conductivity of charge carriers interacting with a strongly coupled gauge theory. Tractable holographic models are probe brane systems, where the charge carriers were represented by
a small number ($N_f$) of probe D-branes with a non-trivial temporal
world volume gauge field (finite charge density) and strongly-coupled gauge field dynamics is encoded in a background geometry. By considering
a large $N_c (\gg N_f)$ gauge theory we ignore back reactions of
probe branes to the background \cite{Karch:2002sh,Karch:2007pd,Hartnoll:2009ns}.

There are three main holographic methods to compute the DC conductivity,
namely,  (1) the (holographic) retarded Green's function method~\cite{Son:2002sd},
(2) the (black hole) membrane paradigm method~\cite{Kovtun:2003wp,Iqbal:2008by}, and
(3) the real-action method~\cite{Karch:2007pd}.

First, the retarded Green's function method is a general holographic method, which can be applied to many transport coefficients.  The boundary to bulk (gravity) Green's function of some field
encodes a retarded correlator of its dual (field theory) operator.
Then, using Kubo formula we can compute transport coefficients
in the hydrodynamic limit. We will not deal with this method so
refer to \cite{Son:2002sd} for more details.

Second, the membrane paradigm method is also general and can be
applied to many transport coefficients.
Especially it enables us to read off the zero momentum limit behavior of the the boundary field theory
from the stretched horizon of the black hole\footnote{The next order finite small momentum behavior such as the diffusion constant requires metric information over all bulk not
only the one at the stretched horizon.  We will come back to this issue in section \ref{sec.7}.}.
As a result, some transport coefficients may be expressed in terms of
combinations of the near-horizon metric components.
This geometric meaning of field theory quantity
elegantly explains universalities
of transport coefficients~\cite{Iqbal:2008by}.
When it comes to DC conductivity,
this method was initially applied to the case with pure metric,
while no charge or background electromagnetic field was presented~\cite{Iqbal:2008by}.
However, it was recently generalized to the system with finite charge
and background electromagnetic field~\cite{Kim:2011qh}.
In this case, in order to use the membrane paradigm method we need to first
{\it geometrize} the background gauge field\footnote{Holographically a conserved charge is encoded in a time component
of the background gauge field ($A_t$)\cite{Kim:2006gp,Kim:2007zm,Kobayashi:2006sb}.}, so that all
physics can be expressed only in terms of an effective metric,
the so-called {\it open string metric} (OSM).
Then the membrane paradigm method can be used in terms of OSM.
It will be explained in more detail in section \ref{sec.3} and \ref{sec.4}.

Third, the real-action method is special in three aspects.
(1) it is tailored for DC conductivity and is not applicable to
other transport coefficients. (2) it applies only to
probe D-brane systems described by the DBI action.
(3) it yields {\it a non-linear
conductivity}
while previous two methods yield {\it a linear
conductivity}
since they are based on linear response theory.
Here, by the linear (non-linear) conductivity, we mean
 an electric field independent (dependent) conductivity\footnote{This is abuse of terminology. Strictly speaking, we should say that ``current'' is linear or non-linear to electric field.}.
These three properties as well as the basic mechanism
of the real-action method stem from
the square root of the DBI action.
The DBI action of Dp-brane is schematically
\begin{equation}
  \int \dd \xi^{p+1} \sqrt{\det(P[G] + \calf)} \,, \nonumber
\end{equation}
where $P[G]$ is the induced metric and $\calf$ is the gauge field strength.
With a finite background electric filed in $\calf$, it turns out
that $P[G] + \calf$ becomes negative at some point, so the DBI action
becomes imaginary. To make the action real
(so the real-action method) we need to turn on
another gauge field compensating the ``negative'' introduced by
an electric field. This compensating gauge field encodes
the current responding to the electric field.
It has been shown, in many examples, that the linear limit of non-linear current obtained by
the real-action method agreed to the results of the retarded
Green's function and the membrane paradigm method.

One of the goals of this paper is to propose the fourth holographic method
to compute DC conductivity based on the OSM and the membrane paradigm.
We will call it the {\it OSM method} for short.
The basic idea was presented in \cite{Kim:2011qh} and applied for the DC conductivity of the D3/D7 brane system, which successfully reproduced the
known results obtained by the real-action method. In this paper, we will
generalize this idea in two ways.
Firstly, we include the Wess-Zumino term, dilaton field,
internal flux field to the formalism in arbitrary dimensions, so that the formalism could be applied to and tested in various models.
However, this method, in its original form, gives us only the linear conductivity since it is based on the linear response theory.
So, secondly, we propose how to obtain a non-linear conductivity in
this OSM-based framework. We believe that the OSM method is
equivalent to the real-action method as far as the final
conductivity formula is concerned. However, it gives us
a new conceptual insight and also a technical simplification.
For example, the condition that the action is real in the real-action method, is interpreted as the condition that the geometry of OSM is regular. As applications of our extended formalism we study D3-D7$^\prime$ \cite{Bergman:2010gm} and D2-D8$^\prime$ systems~\cite{Jokela:2011eb}
showing the fractional and integer quantum Hall current respectively.

Another goal of this paper is the discussion on the strange metal
property. The OSM method as well as the real-action method give us
the conductivity formula expressed in terms of the metric.
This enables us to understand what kinds of bulk geometric properties
allow the strange metallic property.
It is argued \cite{Hartnoll:2009ns} that a pure Lifshitz geometry cannot produce \eqref{eq.1},
based on the general metric-expression of conductivity and dimensional
analysis. However, it was shown \cite{Kim:2010zq} that the light-cone AdS geometry could satisfy \eqref{eq.1}.
We analyze these systems by the OSM method and study the underlying
structure distinguishing them. This will give
us a useful guide for a holographic model building of a strange metal.
We also discuss the effect of
the non-trivial dilaton on the DC conductivity.
Last but not least, we note that the OSM is also useful to compute
other transport coefficients other than the DC conductivity.
As a simple example, we discuss on the charge diffusion at finite
magnetic field and density using OSM.

The paper is organized as follows.
In section 2, we briefly explain what the OSM is.
After presenting a basic idea underlying the OSM method for the DC conductivity in section 3, we show the detailed formalism in section 4.
The formalism are applied to D3-D7$^\prime$~\cite{Bergman:2010gm} and D2-D8$^\prime$~\cite{Jokela:2011eb} model and the light-cone AdS black hole~\cite{Kim:2010zq} in section 5. The first two are related to the quantum Hall effect,
while the last is related to the strange metal property.  In section 6 we revisit
the holographic strange metal property from a general metric point of view, where
the effect of a non-trivial dilaton on the DC conductivity is also briefly
addressed. In section 7, we summarize and discuss how to compute
other transport coefficients from OSM, as well as how to study the
thermodynamics of the effective horizon of OSM.

\section{Open string metric}
\label{sec.2}

In this section we review on the open string metric.
Let us consider probe brane systems, where the DBI and the WZ term determine its dynamics. Schematically
\begin{equation}
  \call = \call_{\mathrm{DBI}} + \call_{\mathrm{WZ}} = \sqrt{-\det(P[G] + \calf )} + P[C] \wedge \calf \cdots  \,, \LA{eq1}
\end{equation}
where $\calf = \tF + 2\pi\a'f$ and P[$\  \ $] denotes pulling back.
Classical configurations of probe branes are determined by the combination of
the pull-backed background fields (metric ($G$), RR field ($C$)) onto the world volume
of the probe branes and the world volume background gauge fields ($\tF$).
Small fluctuations of gauge fields ($f$) can be studied in this background by expanding DBI + WZ in terms of $\a'$.
There may be additional world volume scalar field fluctuations coming from pull-back of metric ($G$) or RR fields ($C$).
In brief, fluctuating fields are interacting with both the induced metric and various background gauge fields.
However, we may understand the dynamics of these fluctuating fields in terms of the {\it effective metric},
that is, {\it open string metric} (OSM), unifying
the induced metric and background gauge fields~\cite{Kim:2011qh}.
In some sense, the background gauge fields are {\it geometrized}.
%
%
Following \cite{Seiberg:1999vs}, we define the (inverse of) the open string metric, $s^{mn}$, and an antisymmetric tensor $\theta^{mn}$ through the simple relation
\begin{align}
\gamma_{mn} &\equiv P[G] + \tF \,,  \\
\gamma^{mn} &=\left(\gamma_{mn}\right)^{-1}=s^{mn}+ \theta^{mn} \, , \label{eq.gamma}
\end{align}
where $s^{mn}$ is the symmetric part and $\theta^{mn}$ is the anti-symmetric part of $\gamma^{-1}$. $s_{mn}$ is defined such that $s_{mn} s^{np}=\delta_m^p$, then it can be shown that
\begin{equation}
s_{mn} =g_{mn}-(\tF g^{-1} \tF)_{mn} \label{eq.sab}  \, ,
\end{equation}
We will refer hereafter to $s_{mn}$ as the open string metric (OSM)\footnote{The fields living in the brane correspond to open string degrees of freedom, hence the name OSM.
}.
Also see \cite{Kim:2011qh} for the discussion from a blackfold \cite{Emparan:2011br} point of view.

A straightforward way to see the appearance of OSM can be
sketched as follows.
If we consider small fluctuations order of $\a'$ around the background
then the DBI action can be expanded in terms of $\a'$ as
\begin{equation}
\begin{split}
  \call_{\mathrm{DBI}} & \sim \sqrt{-\det (\g + 2\pi\a' \g^{(1)} + (2\pi\a')^2 \g^{(2)} + \cdots)} \\
  & \sim \sqrt{-\det\g} \left(1+\half \tr X + \frac{1}{8} (\tr X)^2 - \frac{1}{4} \tr (X^2) + \cdots\right) \,,
\end{split}
\end{equation}
where $X = \g^{-1}(2\pi\a' \g^{(1)} + (2\pi\a')^2 \g^{(2)} +\cdots )$.
Note that the appearance of $\g^{-1}$ in $X$, which is why $\g^{-1}$
plays a role instead of $\g$ in \eqref{eq.gamma}.
For example, let us consider a symmetric second order fluctuation $\g^{(2)}_{mn} \sim \partial_m \varphi \partial_n \varphi $, where $\varphi$ is the
pseudo-scalar meson. This projects out the anti-symmetric
part of $\g^{-1}$ ($\theta^{mn}$) in $\tr X$. So $\tr X \sim \tr \g^{-1} \g^{(2)} \sim  s^{mn}\partial_m \varphi \partial_n \varphi$, which corresponds to the
kinetic term of a pseudo-scalar meson. In a similar way
$\tr X^2 \sim s^{mm'}s^{nn'}f_{mn}f_{m'n'}$ yields a kinetic term of vector mesons. See \eqref{DBI2} and \eqref{DBI27} for more complete expressions,
where note that the anti-symmetric part $\theta^{mn}$ also play a role.

This OSM idea itself was discussed long ago~\cite{Seiberg:1999vs,Gibbons:2000xe}, but its usefulness in the context of gauge/gravity duality applications was proposed recently.
In \cite{Kim:2011qh}, it has been shown
that OSM identifies a new effective event horizon away from a background black hole horizon, when a constant electric field is geometrized.
The position of this horizon agrees to the so-called {\it singular shell}, where the currents due to the electric field are determined by
requiring the action to be real (the real-action method) \cite{Karch:2007pd,
OBannon:2007in,Ammon:2009jt}. (It will be explained in more detail in section \ref{3.2}.)
As a simple application of OSM, the linear DC conductivity of
the D3/D7 system at finite density and magnetic field, was computed
by applying the membrane paradigm~\cite{Iqbal:2008by} to OSM (the OSM method).

\section{Basic ideas for DC conductivity with open string metric}
\label{sec.3}

In this section, we explain the basic idea of
the membrane paradigm method, the real-action method, and the generalized
OSM method in a simple setup, highlighting the essential idea.
A more detailed work will be done in the following section.

\subsection{Linear conductivity: membrane paradigm} \label{123}

We start with a review on how to compute the DC (linear)
conductivity using the membrane paradigm method in the simplest setup - only a Maxwell term in the \emph{diagonal} metric background $g_{mn}$ -
in order to highlight the basic idea \cite{Iqbal:2008by}.
\begin{equation}
  S_{\mathrm{eff}}  =  -  \int \dd^{5}x
  \left[\frac{ \sqrt{-g} }{4\, g^2_{5}} g^{mp}g^{nq} f_{mn} f_{pq} \right]\,,
\end{equation}
where $g_5$ is the 5 dimensional gauge coupling.
The canonical momentum to $a_i$ $(i=x,y,z)$ at fixed radial variable $r$ yields
\be
  {\cal J}^i(r) = - \frac{1}{g_{\mathrm{5}}^2} \sqrt{-g}\, f^{r i}
   \ , \label{GeneralJ}
\ee
The in-falling boundary condition at the horizon ($r=r_H$) implies
\be
f_{ri}(r_H)=  \sqrt{\frac{g_{rr}(r_H)}{-g_{tt}(r_H)}}f_{ti}(r_H) \, , \label{EF1}
\ee
which can be derived by
requiring the fields are functions of only in-going Eddington-Finkelstein coordinate. Thus \eqref{GeneralJ} at the horizon is
\begin{equation} \label{kk1}
{\cal J}^i(r_H)= \left. \frac{1}{g_{5}^2}\sqrt{\frac{g}{g_{tt}g_{rr}}}g^{ii}f_{it} \right|_{r=r_H}\, ,
\end{equation}
with \eqref{EF1}.
However, in the zero momentum limit,
\be
\partial_r {\cal J}^i(r)|_{\omega\rightarrow 0} = \partial_r f_{it}(r)|_{\omega\rightarrow 0} = 0 \, , \label{check1}
\ee
which can be shown from the Maxwell equations. Thus, there is no
flow of \eqref{kk1} in $r$ and it is valid at any $r$,
especially at $r=\infty$.
Recall, from AdS/CFT dictionary, that the expectation value ($j^\mu$) of the conserved current
in the boundary field theory is identified with
${\cal J}^i(r \rightarrow \infty)$ and the conductivity tensor ($\sigma^{ij}$) can be written as
\be
  j^i(k^\mu) =
     {\cal J}^i(r \rightarrow \infty)(k_\mu) \equiv \sigma^{ij}(k_\mu) f_{jt}(r \rightarrow \infty) \, .
\ee
As a result, the DC conductivity is
\begin{equation}
   \s^{ii}(k=0) = \left.\frac{1}{g_{5}^2}\sqrt{\frac{g}{g_{tt}g_{rr}}}g^{ii}\right|_{r=r_H} \,.
\end{equation}

The relation of this general argument to OSM is straightforward.
OSM is a metric, so we can simply replace $g_{mn} \ra s_{mn}$.
However, OSM is often not a diagonal metric and the starting action may have an extra term other than a Maxwell term due to anti-symmetric $\theta^{mn}$.
So the formalism should be extended to a Maxwell plus topological term in \emph{non-diagonal} or \emph{non-static, stationary} metrics.
These have been done in \cite{Kim:2011qh} and the results will be reproduced or shown in the following section \ref{sec.4}.

\subsection{Non-linear conductivity}
\label{3.2}

The membrane paradigm is based on linear response theory so it
describes a linear conductivity, which means that
the conductivity is independent of the electric field.
However, non-linear conductivity, which depends on the electric field,
has been obtained in many holographic systems from
the real-action condition. Let us review this method in a simple setup,
a supersymmetric D3/D5 system with a finite electric field ($\tE$) and the
corresponding current ($J_x$).
It is effectively described by a one-dimensional action~\footnote{See
\eqref{gen2p1} for a more general expression and the derivation.
The action in principle is a functional of $A_x'$ and $J_x$ is
the conjugate variable to $A_x'$. We simply replaced $A_x'$ by $J_x$ in
\eqref{act1} to discuss real-action condition,
but it also could be discussed in the Legendre transformed action \eqref{act2},
where $J_x$ is an independent variable.}
\begin{equation} \label{act1}
  S \sim  - \int_{r_H}^{\infty} \dd r g_{\Omega\Omega}^2
  \sqrt{-g_{tt} g_{rr} g_{xx}^2} \sqrt{\frac{\xi}{\chi}} \,,
\end{equation}
where
\begin{equation}
  \xi = -g_{tt} g_{xx}^2 - g_{xx} \tE^2 \,, \quad
  \chi = -g_{tt} g_{xx}^2 g_{\Omega\Omega}^2 - g_{xx} J_x^2 \,.
\end{equation}
On the other hand, the Legendre transformed action reads
\begin{equation}\label{act2}
  S_{\mathrm{LT}} \sim  - \int_{r_H}^{\infty} \dd r \frac{ \sqrt{g_{rr}} }{\sqrt{-g_{tt}}g_{xx}}  \sqrt{\xi \chi} \,.
\end{equation}
The current $J_x$ is determined by {\it real-action condition}.
The idea comes from the fact that $\xi$ becomes negative near the
horizon $g_{tt} \ra 0$, whatever $\tE$ is.
We call the location where the sign of $\xi$ flips {\it singular shell}.
However, $\chi$ would be always positive if we did not introduce $J_x$.
So by introducing and adjusting $J_x$ to flip the sign $\chi$ at the singular shell, we may keep the action real.
i.e.
\begin{align}
  \chi(r_s) &= 0  \qquad  \Longleftrightarrow \label{Jx1} \\
  J_x &= \left.\sqrt{-g_{tt} g_{xx} g_{\Omega\Omega}^2}\right|_{r=r_s(\tE, r_H)} = g_{\Omega\Omega}(r_s)\tE \equiv \s(r_s) \tE  \,, \label{Jx111}
\end{align}
where $r_s$ is defined such that $\xi(r_s)=0$, so $(-g_{tt}g_{xx}) \ra \tE^2$ in the second line.
The conductivity $\s(r_s)$ is a function of $\tE$ and $r_H$ via $r_s(\tE,r_H)$, so
the current is nonlinear in $\tE$.

We may understand this real-action condition from a geometric point of view
of OSM. If we introduce only $\tE$ then the geometry of $s_{mn}$
becomes singular at the singular shell.
It can be seen from the Ricci scalar ($\calr$) near $r_s$ ($\xi \sim 0 $)
\begin{equation}
  \calr \sim \frac{(g_{xx} g_{tt}' + g_{tt} g_{xx}')^2}{2 \xi^2 g_{rr}} \,.
\end{equation}
To make the geometry regular we can introduce the current $J_x$ then
it changes OSM and yields
\begin{equation}
   \calr \sim \frac{\chi (g_{xx} g_{tt}' + g_{tt} g_{xx}')^2}{2 \xi^2 g_{rr}
   g_{tt}g_{xx}g_{\Omega\Omega}^2} \,,
\end{equation}
of which regularity condition gives us the same current as \eqref{Jx1}.

In principle, we may be able to derive currents from this {\it regular
geometry condition}. However, in practice, there is a more efficient way,
which is based on two observations.

First, note that the currents do not affect the singular shell position and
the currents are computed after the singular shell is given. It means that
we can first identify the singular shell without introducing
the currents by focusing on the singularity of our OSM.
In general, the singular shell condition was shown to be \cite{OBannon:2007in,Ammon:2009jt}
\begin{equation} \label{xx}
  \xi(r_s)=\det \g_{\m\n} (r_s) = 0 \,,
\end{equation}
where $\m,\n$ are indices only for the field theory directions.
At the point $r=r_s$, the geometry of OSM is also singular without the currents\footnote{It turns out that $\xi=0$ yields $\det s = 0$.}.

Second, we note that the linear limit of conductivity is simply
$\sigma(\tE=0)$ in \eqref{Jx111} as shown in \cite{Kim:2011qh}. All non-linearity
comes in only from the singular shell position $r=r_s(\tE, r_H)$.
This was also noted in \cite{Mas:2009wf}.

Thus we propose the following way to compute non-linear DC conductivity:
(1) compute the linear conductivity using OSM and membrane paradigm,
(2) compute the singular shell position $r_s$ from \eqref{xx} with finite $\tE$ (and $\tB$),
(3) apply the same formula obtained in (1) at $r = r_s$ instead of $r=r_H$,
which is consistent with the fact that the singular shell is
a real effective horizon, instead of a background black hole horizon \cite{Kim:2011qh}. In the following sections,
we will show this process reproduces the same results as
the real-action method.

\subsection{Minkowski embedding: no world volume horizon}

For Minkowski embedding, we cannot apply the real-action method since
there is no singular shell on the world volume. Therefore there is no reason to introduce currents, for example, $J_x$ in \eqref{act1}\footnote{However
there is a subtlety at zero temperature. It was shown \cite{Evans:2011tk}
that we have to introduce the Hall current (no Ohmic current)
even without a singular shell
at zero temperature and $E < B$ to make the action real.}.
Also from the OSM point of view,
the geometry is regular everywhere and there seems to be no reason to
introduce the current. However, we still should require
regularity on the gauge field configuration in the IR, which does not allow the Ohmic current but the Hall current.
We also expect that the Hall current be linear
since all non-linearity comes only from the singular shell.
This requirement was proposed in \cite{Bergman:2010gm} and we will
show how to apply it in our OSM context in the following section.

\section{DC conductivities in terms of open string metric}
\label{sec.4}

In this section we derive the conductivity formula in 2+1 dimensions and
3+1 dimensions by the OSM method. Especially we generalized \cite{Kim:2011qh} by including the Wess-Zumino term, dilaton field as well as internal flux field. Even though the logic of 2+1 dimensions and 3+1 dimensions is the same, we do it separately, since the 2+1 formalism
is too simple to be extended to higher dimensions, while 3+1 dimensions
formula shows all general features to be extended to dimensions higher than
3+1.

\subsection{2+1 dimensions}

Let us consider a Dq-brane system in higher dimension ($D \ge 5$), sharing $t,x,y$ field theory space.
The induced metric and background gauge fields are assumed to be
\begin{align}
\dd s^{2}_{q} & =g_{tt} \ud t^{2} +\sum\limits^{2}_{i=1} g_{ii} \ud x^{2}_{i} + g_{rr} \ud r^{2} + \dd s_{\nnn}^2 \, ,  \label{2p1} \\
2 \pi \alpha' A & \equiv  \tA + 2\pi \alpha' a
  \equiv \tilde{A}_{t}(u)\, \dd t  + \tilde{B} \, x\, \dd y
 + 2\pi \alpha' a  \, ,
\end{align}
where $I=q-3$ denotes the dimension of internal space.
$\dd s_{\nnn}^2$ is the metric of the internal
space. If there is a non-trivial flux in the internal space to stabilize the embedding, a specific parametrization is favored.
See, for example, \eqref{7p3} and \eqref{8p3}.
The probe brane embedding information, which is assumed to be
a function of $r$, is hidden in the induced metric, while
$\tA_t$ encodes a finite density and $\tB$ a magnetic field.
There may be specific RR fields ($C_n$) depending on the concrete background.
For example, (\ref{2p1}) includes the supersymmetric D3/D5($I=2$, $C_4$) \cite{Jensen:2010ga,Evans:2010hi,Evans:2011tk},
non-supersymmetric D3/D7$^\prime$($I=4$, $C_4$) \cite{Davis:2008nv, Bergman:2010gm},
D2/D8$^\prime$($I=5$, $C_3, C_5$) system \cite{Jokela:2011eb},
or the probe branes in Lifshitz geometry \cite{Hartnoll:2009ns}.
Since the WZ term should be dealt with case by case, we start with the DBI action ignoring
the WZ term for a moment.

\paragraph{DBI term} We assume that the matrix $\g = g + 2\pi\a' F$ is a direct sum of
the submatrix in the space of $m = 0,1,2,r$ and the internal space
$\a = 4, \cdots , q+1$\footnote{We summarize our notations and conventions: Background space time: $M,N$, The holographic radial coordinate: $r$,
internal sphere: $\a,\b$,
Field theory spacetime: $\mu,\nu, = 0,1,2$, only space: $i,j = 1,2$,
Field theory spacetime plus $r$: $m,n = 0,1,2,r$.
The variables with tilde include $2\pi\a'$. }.
We also assume that the internal gauge field is a function of the
internal space.
Thus, $\det \g = \det \g_{mn} \det \g_{\a\b}$,
where
\begin{equation} \label{Theta1}
  \det \g_{\a\b} \sim \Theta(r) \times \mathrm{a\ function\ of}\ \xi^\a \,,
\end{equation}
which makes the integration over the internal space to be done separately.
%
%
\begin{equation} \label{x123}
\begin{split}
S_{\mathrm{DBI}}&=-N_{f}T_{Dq}\int \dd^{q+1}\xi e^{-\phi}\sqrt{-{\rm det}\g} \\
&=-N_{f}T_{Dq}V_{(I)} \int \dd t \dd\vec{x} \dd r e^{-\phi}\sqrt{\Theta}\sqrt{-{\rm det}\gamma_{mn}} \\
& \equiv  \mathcal{N}\int \dd t \dd\vec{x} \dd r \call_{\mathrm{DBI}} \,  ,
\end{split}
\end{equation}
where $V_{(I)}$ is the result of the integration over the internal space,
of which simplest case is the volume of the internal sphere.
The normalization constant absorbs $V_{(I)}$ and is defined as
\begin{align} \label{Ns}
  \caln &\equiv N_{f}T_{Dq}V_{(I)} \,, \qquad
  \caln' \equiv (2\pi\a')^2 \caln \,,
\end{align}
where $\caln'$ is defined for later convenience.

The leading order Lagrangian reads
\begin{eqnarray} \label{GenDBI}
  \call_{\mathrm{DBI}}^{(0)} = - e^{-\phi} \sqrt{\Theta \k} \sqrt{-g_{tt} g_{rr} -\tA_t'^2 } \,,
\end{eqnarray}
where $\k$ is defined by the determinant of the matrix of $\g_{ij}$ only for $i,j=1,2$.
\begin{eqnarray}
  \k \equiv \det \g_{ij} = \tB^2 + g_{xx} g_{yy} \,.
\end{eqnarray}

Since $\tA_t$ is a cyclic quantity, it is convenient to trade it with a conjugate conserved quantity, $\hJ_t$, defined by
\begin{equation}
  \hJ_t \equiv \frac{\partial \call_\mathrm{DBI}^{(0)}}{\partial \tA_t'} =
  \frac{e^{-\phi}  \tA_t' \Th \k }{\sqrt{-(\tA_t'^2 + g_{rr} g_{tt}) \,
  \Th \k \, }} \, , \label{jt1}
\end{equation}
which yields
\begin{eqnarray} \label{Atp1}
  \tA_t' = \sqrt{ - \frac{\hJ_t^2  g_{rr} g_{tt}}{\hJ_t^2  + e^{-2\phi}\Th \k}} \ .
\end{eqnarray}

The sub-leading action reads:\footnote{In general, there
are other modes such as scalar modes and gauge fields in the internal space, which are coupled each other.
For a complete analysis of fluctuations
we have to consider them all together.
However, since these extra modes are decoupled for the study of DC conductivity as shown in the appendix of \cite{Kim:2011qh}, we
omit them here. \label{mmm}}
\begin{equation} \label{DBI2}
S_{\rm DBI}^{(2)}=-\mathcal{N}^{\prime}\int \dd t \dd\vec{x} \dd r \, \left[\frac{\sqrt{-s}}{4g^{2}_{4}}
s^{mp}s^{nq}f_{mn}f_{pq}+\frac{1}{8}\epsilon^{mnpq}f_{mn}f_{pq}Q\right]\,,
\end{equation}
where $ \mathcal{N}^{\prime}=(2\pi\alpha^{\prime})^{2}\mathcal{N}$ is
a normalization factor including $\a'^2$, $g^{2}_{4}=\frac{\sqrt{-s}}{e^{-\phi} \sqrt{-\det\g_{mn}} \sqrt{\Theta}}$ is an effective $r$-dependent
coupling, and OSM \eqref{eq.sab} are written as
\begin{equation} \label{OSM1}
\begin{split}
&s_{mn}\dd x^{m}\dd x^{n} = \left[g_{mn}-(\tF g^{-1} \tF)_{mn}\right] \dd x^{m}\dd x^{n} \\
& \qquad \qquad \quad \, = g_{tt}\mathcal{G}^{2}\dd t^{2}+g_{rr}\mathcal{G}^{2}\dd r^{2}
+\frac{\kappa}{g_{yy}}\dd x^{2}+\frac{\kappa}{g_{xx}}\dd y^{2}
\,, \\
&\mathcal{G}^{2}=\frac{e^{-2\phi} \Th \kappa}
{\tilde{J}^{2}_{t}+e^{-2\phi}  \Th \kappa} \,, \quad
 Q =-\frac{e^{-\phi}\sqrt{-\det\gamma_{mn}}\sqrt{\Th}}{8}\epsilon_{mnpq}\theta^{mn}\theta^{pq}
= - \frac{\tB \hJ_t}{\kappa} \,,
\end{split}
\end{equation}
with $\e_{txyr} = 1$ and
non-vanishing  $\theta^{mn}$ components
\begin{equation} \label{theta1}
\theta^{tr}= \frac{\tilde{A}^{\prime}_{t}}{\tilde{A}^{\prime2}_{t}+g_{tt}g_{rr}} = -\frac{e^{\phi}\hJ_t}{\sqrt{-\det\gamma_{mn}}\sqrt{\Th}} \,, \qquad
\theta^{xy}=-\frac{\tilde{B}}{\k}\,,
\end{equation}
Note that the effects of density ($\hJ_t$) and magnetic field ($\tB$) are geometrized through $\calg$ and $\k$.

\paragraph{Wess-Zumino term} The relevant Wess-Zumino term to our discussion is
\begin{equation} \label{GenWZ}
    S_{\mathrm{WZ}} = \frac{N_f T_{Dq} (2\pi\a')^2}{2} \int P[C_{q-3}]\wedge F \wedge F \, ,
\end{equation}
This term appears in a non-supersymmetric D3/D7($C_4$) system \cite{Bergman:2010gm}
or D2/D8($C_5$) system \cite{Jokela:2011eb}.

The leading order action \eqref{GenWZ} is
\begin{equation} \label{uuuu}
    S_{\mathrm{WZ}}^{(0)} = \caln \int \dd t \dd\vec{x} \dd r \, C_{q-3} \tF_{0r}  \tF_{12}  = \mathcal{N}\int \dd t \dd\vec{x} \dd r \call_{\mathrm{WZ}}^{(0)}  \, ,
\end{equation}
This will modify \eqref{jt1} as
\begin{equation} \label{jt2}
\begin{split}
  \hJ_t = \frac{\partial \big[\call_\mathrm{DBI}^{(0)}+\call_\mathrm{WZ}^{(0)} \big] }{\partial \tA_t'} &=
  \frac{e^{-\phi}  \tA_t' \Th \k  }{\sqrt{-(\tA_t'^2 + g_{rr} g_{tt}) \, \Th \k \,  }} - C_{q-3}(r) \tB  \\
  & \equiv \bar{J}_t(r)-  C_{q-3}(r) \tB  \,.
\end{split}
\end{equation}
Then, we simply need to replace
\begin{equation} \label{newJt}
  \hJ_t \ra \bar{J}_t(r) =\hJ_t + C_{q-3}(r)\tB \,,
\end{equation}
in the subsequent equations \eqref{OSM1} and \eqref{theta1}.

Note that a conserved quantity $\hJ_t$ is a constant, but $\bar{J}_t(r)$
is a function of $r$, whose $r$ dependence will be compensated by
$C_{q-3}(r)$ to render $\hJ_t$ to be constant in all $r$. There are
two different contributions to a total charge $\hJ_t$:
topological charge and strings. The former is expressed by
a nontrivial function of $r$, $C_{q-3}(r)$,
while the latter is essentially a delta-function source
at the IR end of the probe brane (say $r_0$). Its existence is
manifested by the boundary condition of nonzero $A_t'(r_0)$.
Thus in the case of vanishing string source (Minkowski embedding),
we require $A_t'(r_0) = 0$, which yields
\begin{equation} \label{Min}
  \hJ_t = - C_{q-3}(r_0) \tB \,,
\end{equation}
from \eqref{jt2} \cite{Bergman:2010gm}.
This relates the charge density and the magnetic field through a topological (so discrete) internal flux $C_{q-3}(r_0)$, showing a typical property of a quantum Hall state.

At sub-leading order, \eqref{GenWZ} contribute as quadratic fluctuations terms, so in addition to \eqref{DBI2} we have
\begin{eqnarray} \label{WZ2}
  S_{\mathrm{WZ}}^{(2)} = \caln' \int \dd \xi^4 C_{q-3} \epsilon^{ji} f_{j0} f_{ri} + \cdots \,,
\end{eqnarray}
where we have explicitly shown only the terms which are relevant to the DC conductivity.

\paragraph{Membrane paradigm} Now we follow the logic of the membrane paradigm method presented in section \ref{123} with a generalized action \eqref{DBI2} plus \eqref{WZ2}.
The canonical momentum of $a_i$ from \eqref{DBI2} and \eqref{WZ2} is
\begin{equation} \label{CanMom}
\mathcal{J}^{i}(r)= -\frac{\mathcal{N}^{\prime}}{g^{2}_{4}}\sqrt{-s}f^{ri}
-  \mathcal{N}^{\prime} Q\epsilon^{ji}f_{j0}  + \caln' C_{q-3} \e^{ji} f_{j0}  ,
\end{equation}
and the conductivity tensor ($\s^{ij}$) reads, by AdS/CFT dictionary,
\begin{equation}
j^{i}(k^{\mu}) \equiv \mathcal{J}^{i}(r\rightarrow \infty)(k_{\mu})
\equiv\sigma^{ij}(k_{\mu})f_{jt}(r\rightarrow \infty) =\sigma^{ij}(k_{\mu})\cale_{j} .
\end{equation}
At $k^\mu \ra 0$ limit, $\calj^i$ and $f_{jt}$ are constant in $r$, so we may evaluate it in any IR coordinate (say $r_0$):
\begin{equation} \label{ji2D}
j^i =  \caln' \left[-\frac{1}{g^{2}_{4}}\sqrt{-s} s^{rr} s^{ij} f_{rj}
-  Q \epsilon^{ij} f_{jt}  -  C_{q-3} \e^{ij} f_{jt} \right]_{r\rightarrow r_0} \,.
\end{equation}
For a black hole embedding we evaluate it at the stretched horizon and
make use of a regularity condition at the horizon,
\begin{equation} \label{reg1}
f_{rj}= \sqrt{\frac{s_{rr}}{-s_{tt}}}f_{tj} .
\end{equation}
From the Ohm's law $j^{i} =\s^{ij} f_{jt} =  \s^{ij} \cale_j $ we have the conductivity
\begin{equation} \label{sij2D}
\s^{ij} =  \caln' \left[\frac{1}{g^{2}_{4}}\sqrt{\frac{s}{s_{tt}s_{rr}}}s^{ij}
-   Q \epsilon^{ij}  -  C_{q-3} \e^{ij}  \right]_{r\rightarrow r_H} \,.
\end{equation}
This is a conductivity in the limit of a small electric field compared to any other scale, which is electric field independent (a linear conductivity).

To compute  a non-linear conductivity we follow the proposal
in section \ref{3.2}.  First we find the effective horizon $r_s$
by the condition \eqref{xx}:
\begin{equation}
  \det \g_{\m\n} = \left[\tilde{B}^{2}g_{tt}+\tilde{E}^{2}_{x}g_{yy}+\tilde{E}^{2}_{y}g_{xx}+g_{tt}g_{xx}g_{yy}\right]_{r \ra r_s} = 0 \,.
\end{equation}
For example, in the case of supersymmetric D3/D7(D5) \cite{OBannon:2007in,Evans:2011tk} case it gives us
\begin{equation}
  r_s =  \half \left(T^4 -\tB^2 + \tE^2 + \sqrt{-4\tB^2 \tE^2 + (T^4 + \tB^2 + \tE^2)^2}\right)^{1/4} \, .
\end{equation}
Note that if $\tE = \tB = 0$, then $r_s  = T$ as expected.
In general $r_s > T$ always holds and finite electric field always
increases the effective horizon and effective temperature.
We evaluate the conductivity at $r=r_s$
\begin{equation} \label{sij2D1}
\s^{ij} =  \caln' \left[\frac{1}{g^{2}_{4}}\sqrt{\frac{s}{s_{tt}s_{rr}}}s^{ij}
-   Q \epsilon^{ij}  -  C_{q-3} \e^{ij}  \right]_{r\rightarrow r_s} \,,
\end{equation}
where
$\tE_i$ dependence enters only through the singular shell position, $r_s$.  As $\tE_i \ra 0$, $r_s$ goes to $r_H$, then we can recover the linear conductivity.

For the Minkowski embedding we evaluate it at the IR end of the embedding (say $r_0$). Since there is no sink such as a black hole horizon and a source such as a charge from a string, gauge fields have to be regular
at $r=r_0$.
This sets $f^{ri}(r_0) = 0$ in \eqref{CanMom}. Furthermore we have $\tA_t'(r_0)=0$
in \eqref{jt2} so $Q(r_0)=0$.
Therefore the conductivity comes only from the WZ term in \eqref{ji2D}
\begin{equation} \label{sHall}
  \s^{ij} = - \caln' C_{q-3}(r_0) \e^{ij} \,
  =  \caln' \frac{\hJ_t}{\tB} \e^{ij} \,,
\end{equation}
where we used \eqref{Min}.
Note that the Ohmic conductivity is zero but the Hall current is non-zero, which is a standard steady state Hall effect result at large magnetic field.
The Hall conductivity may be classical (continuous) or quantum.
In this holographic context, \eqref{sHall} corresponds to (integer or
fractional) quantum Hall conductivity while the second term in
\eqref{sij2D1} is classical. Thus, the quantum nature of the
conductivity is topological from a holographic point of view, which comes from the internal flux quantization, yielding discrete values of $C_{q-3}(r_0)$ \cite{Bergman:2010gm,Jokela:2011eb}.

\subsection{3+1 dimensions}

The logic of 3+1 dimensions is the same as that of 2+1 dimensions, so we will
be in brief, only highlighting the differences from 2+1 dimensional case.

Let us consider a Dq-brane system in higher dimension ($D \ge 6$), sharing $t,x,y,z$ field theory space.
The induced metric and background gauge fields are assumed to be
\begin{align} \label{3p1}
\dd s_q^{2} &=g_{tt} \dd t^{2} +\sum\limits^{3}_{i=1} g_{ii} \dd x^{2}_{i} + g_{rr} \dd r^{2} +  \dd s^2_{(I)}  \,, \\
 2 \pi \alpha' A &\equiv  \tA + 2\pi \alpha' a
\equiv \tilde{A}_{t}(u)\, \dd t + \tB_y\, z\, \dd x  + \tilde{B}_{z}\, x\, \dd y+  \tilde{B}_{x}\, y\, \dd z
 + 2\pi \alpha' a \,,
\end{align}
where $I=q-4$ denoting the dimension of internal space.
For an isotropic metric, we may start with only two $\tB_i$ field components:
one is parallel to the electric filed and the other is orthogonal to it.
However, we keep all three components to consider more general cases
(without $SO(3)$ symmetry).
There may be specific RR fields ($C_n$) depending on the concrete background.
For example, (\ref{3p1}) includes the supersymmetric
D3/D7($n=3$, $C_4$)~\cite{Karch:2007pd,OBannon:2007in,Ammon:2009jt,Mas:2008qs,Evans:2010iy,Evans:2011mu,Evans:2011tk},
D4/D8($n=5$, $C_3$)~\cite{Bergman:2008sg,Kim:2008zn} system,
or the probe branes in Lifshitz geometry~\cite{Hartnoll:2009ns}.

\paragraph{DBI term} We start with the DBI action \eqref{x123}, where $m,n=0,1,2,3,r$.
The leading order Lagrangian is of the same form as the 2+1 dimensional case \eqref{GenDBI}
\begin{eqnarray} \label{GenDBI1}
  \call_{\mathrm{DBI}}^{(0)} = - e^{-\phi} \sqrt{\Theta \k} \sqrt{-g_{tt} g_{rr} -\tA_t'^2 } \,,
\end{eqnarray}
with a different $\k$ defined by the determinant of the sub-matrix of $\g_{ij}$ only for $i,j=1,2,3$.
\begin{eqnarray} \label{ka3}
  \k \equiv \mathrm{det}\g_{ij}  = g_{xx}g_{yy}g_{zz} +  \sum_{i=1}^3 g_{ii}^2 B_i^2 \ .
\end{eqnarray}
Consequently, $\tA'_t$ yields the same form as the 2+1 dimensional case (with different $\k$)
\begin{eqnarray}
  \tA_t' = \sqrt{ - \frac{\hJ_t^2  g_{rr} g_{tt}}{\hJ_t^2  + e^{-2\phi}\k \Th }} \ . \nn
\end{eqnarray}
The sub-leading action reads in terms of OSM:%
\begin{equation} \label{DBI27}
S_{\rm DBI}^{(2)}=-\mathcal{N}^{\prime}  \int \dd t \dd\vec{x} \dd r e^{-\phi}\left[\frac{\sqrt{-s}}{4g^{2}_{5}}
s^{mp}s^{nq}f_{mn}f_{pq}+\frac{1}{8}\epsilon^{mnpql}f_{mn}f_{pq}Q_{l}\right] \,,
\end{equation}
where $ \mathcal{N}^{\prime}=(2\pi\alpha^{\prime})^{2}\mathcal{N}$, $g^{2}_{5}=\frac{\sqrt{-s}}{e^{-\phi} \sqrt{-\det\g_{mn}} \sqrt{\Theta}}$,
and
\begin{equation}
\begin{split}
& s_{mn}\dd x^{m} \dd x^{n}  = \left[g_{mn}-(\tF g^{-1} \tF)_{mn}\right] \dd x^{m}\dd x^{n} \\
& \qquad \qquad \quad \, = g_{tt}\mathcal{G}^{2}\dd t^{2}+g_{rr}\mathcal{G}^{2}\dd r^{2}
 +\frac{\kappa \d^{ij}g_{ij} - B_{i}B_{j}g_{ii}g_{jj}}{g_{xx}g_{yy}g_{zz}}\dd x^i \dd x^j
+g_{\Omega\Omega}d\Omega^{2}_{n},  \\
& \mathcal{G}^{2}  =\frac{e^{-2\phi}\Theta\kappa}
{\tilde{J}^{2}_{t}+e^{-2\phi}\Th\kappa} \ , \quad
 Q_l  =-\frac{e^{-\phi}\sqrt{-\det\gamma_{mn}}\sqrt{\Th}}{8}\epsilon_{mnpql}\theta^{mn}\theta^{pq} = \frac{\tB_l g_{ll} \hJ_t }{\k},
\end{split}
\end{equation}
with $\e_{txyzr} = 1$ and
the non-vanishing  $\theta^{mn}$ components are given by
\begin{equation} \label{theta17}
\theta^{tr}= \frac{\tilde{A}^{\prime}_{t}}{\tilde{A}^{\prime2}_{t}+g_{tt}g_{rr}}= -\frac{e^{\phi}\hJ_t}{\sqrt{-\det\gamma_{mn}}\sqrt{\Th}} \,, \qquad
\theta^{ij}=-\frac{\e^{ijk}\tilde{B}_k G_{kk}}{\k},
\end{equation}
Note that the effects of density ($\hJ_t$) and magnetic field ($\tB$) are geometrized through $\calg$ and $\k$.
The Wess-Zumino term can be considered in a similar way as the 2+1 dimensional case, so we omit it.

\paragraph{Membrane paradigm} The canonical momentum of $a_i$ from \eqref{DBI27} is
\begin{equation}
\mathcal{J}^{i}(r)= -\frac{\mathcal{N}^{\prime}}{g^{2}_{5}}\sqrt{-s}f^{ri}
- \mathcal{N}^{\prime}\epsilon^{jik}f_{j0}Q_{k},
\end{equation}
%
which yields the conductivity (with a regularity condition at the horizon \eqref{reg1})
\begin{equation}\label{LCcond}
\begin{split}
\sigma^{ii} &=  {\cal N}' \left. \frac{1}{g_{5}^2}\sqrt{\frac{s}{s_{rr}s_{tt}}} \frac{1}{s_{ii}}   \right|_{r\to r_s} \,,\\
\sigma^{ij} &= \left. -{\cal N}' Q_k \epsilon^{kij} \right|_{r\to r_s} \,,
\end{split}
\end{equation}
For an off-diagonal metric in time and space,
\be\label{eq.nondiagmet}
\dd s^2=s_{tt} \dd t^2  +s_{rr}\dd r^2 + s_{xx} \dd x^2+2s_{tx}\dd t \dd x
+ s_{yy}\dd y^2 + s_{zz}\dd z^2  \, ,
\ee
The conductivity is generalized to \cite{Kim:2011qh}
\begin{equation}\label{LCcond11}
\begin{split}
\sigma^{ii} &=  {\cal N}' \left. \frac{1}{g_{5}^2} \frac{\sqrt{-s}}{\sqrt{s_{rr}}\sqrt{-s_{tt}s_{xx}+s_{tx}^2}}\frac{\sqrt{s_{xx}}}{s_{ii}}   \right|_{r\to r_s} \,,\\
\sigma^{ij} &= -\left. {\cal N}' Q_k \epsilon^{kij}\right|_{r\to r_s} \,,
\end{split}
\end{equation}
which reduces to \eqref{LCcond} when $s_{tx} = 0$.

Finally note that the OSM method mainly involves the matrix operations.
The final conductivity expressions are expressed in terms of
$s^{mn}$ and $\theta^{mn}$, so to get them we need to
evaluate the inverse of $g_{mn}$ and some multiplications with $\tF$ and summations of matrices \eqref{eq.sab}. However, these operations can be
easily done by Mathematica or Maple. A technical simplification
of the OSM method is that we don't need to try to make a
``good combination'' to apply the real-action method.
For example, see \eqref{ttt11}-\eqref{ttt13}.
To apply the real-action method, we have to arrange the terms
in the action in the way written in \eqref{ttt11}, by which
we mean ``good combination'', with
the definition \eqref{ttt12}. Then we have to make sure
\eqref{ttt13} really means the real-action condition.
These are non-trivial from scratch. However, the OSM method
only asks some matrix multiplications.

\section{Applications}

In this section we apply the general formula \eqref{sij2D1} and \eqref{sHall} to the D3-D7$^\prime$ \cite{Bergman:2010gm}, D2-D8$^\prime$ \cite{Jokela:2011eb}; and \eqref{LCcond11}
to the Light-cone AdS system \cite{Kim:2010zq}.  The first two models
are holographic reformulations of the fractional/integer quantum Hall effect, where
the quantum Hall states are represented by Minkowski embedding of the probe branes
and the metallic states are represented by black hole embedding.  The last model
describes universal features of strange metals using ${\rm Schwarzschild-AdS_{5}}$
black holes in light-cone coordinates.  Since our main purpose in this section
is a formalism rather than phenomena, we will not go into details of interesting phenomenological
implications of the models, for which we refer to the original papers.
We will be brief only highlighting how the OSM method works.
All our results by the OSM method agree to those in the original papers where
the real-action method was used. Note that these
are non-trivial consistency checks of different holographic methods.

\subsection{D3-D7$^\prime$ and D2-D8$^\prime$: Hall current}

\paragraph{D3-D7$^\prime$}

This model was proposed in \cite{Rey:2008zz,Myers:2008me,Bergman:2010gm}, where the D3-D7$^\prime$ brane configuration is given as
\begin{equation}
\label{D3D7p}
\begin{tabular}{ccccccccccc}
& 0 & 1 & 2 & 3 & 4 & 5 & 6 & 7 & 8 & 9 \\
D3 & $\bullet$ & $\bullet$ & $\bullet$ & $\bullet$ & & & & &  \\
D7 & $\bullet$ & $\bullet$ & $\bullet$ &  & $\bullet$ & $\bullet$ & $\bullet$
&  $\bullet$ & $\bullet$ &  \\
\end{tabular}
\end{equation}
This configuration is non-supersymmetric and unstable because the branes are
repelled from one another in the $x^{9}$ direction. To ensure the stability,
we assume that D7-brane wraps $S^{2}\times S^{2}$ inside $S^{5}$ and we introduce
the following magnetic fluxes on the two $S^{2}$s:
\begin{equation}
2\pi\a'F=\frac{1}{2}\left(f_{1}\dd \Omega^{(1)}_{2}+f_{2}\dd \Omega^{(2)}_{2}\right)\,, \qquad f_{i}=2\pi\alpha^{\prime}n_{i}\,, \label{7p1}
\end{equation}
where $\dd \Omega_2^{(i)} \equiv \sin \theta_{i} \dd \theta_{i} \wedge \dd \phi_{i}$ and $n_i$ are integers. For a detailed argument
for this stabilization, we refer to \cite{Bergman:2010gm}.
In addition, we consider the gauge fields
\begin{equation}
2\pi\a'A=\tilde{A}_{t}\dd t+\tilde{B}x \, \dd y \,, \label{7p2}
\end{equation}
representing the charge density and the magnetic field.
With an assumption that the scalars $z(=x^3)$ and $\psi(=x^9)$ are
functions of only the radial coordinate, $r$,
the pull-back of the metric and the RR 4-from field on the probe D7-brane reads
\begin{equation} \label{7p3}
\begin{split}
\dd s^{2}_{D7}&=r^{2}(-f(r)\dd t^{2}+\dd x^{2}+\dd y^{2})+\left(\frac{1}{r^{2}f(r)}+r^{2}z^{\prime}(r)^{2}
+\psi^{\prime}(r)^{2}\right)\dd r^{2} \\
& \quad +\cos^{2}\psi (\dd \Omega^{(1)}_{2})^{2}+\sin^{2}\psi(\dd\Omega^{(2)}_{2})^{2}\,, \\
C_{4}&=r^{4}dt\wedge\ dx \wedge dy \wedge dr+\frac{1}{2}c(r)d\Omega^{(1)}_{2}\wedge d\Omega^{(2)}_{2}\,,
\end{split}
\end{equation}
where $f(r)=1-r_{H}^{4}/r^{4}$, $(\dd \Omega_2^{(i)})^2 \equiv
\dd \theta_i^2 + \sin^2 \theta_i \dd \phi_i $ and
\begin{equation}
\begin{split}
c(r) & \equiv c(\psi(r)) = \frac{1}{8\pi^{2}}\int_{S^{2}\times S^{2}}C_{4}\\
& =\psi(r)-\frac{1}{4}\sin4\psi(r)- \psi(\infty)-\frac{1}{4}\sin4\psi(\infty)\,.
\end{split}
\end{equation}
The gauge freedom of the RR 4-from field is fixed by requiring $c(r\ra\infty) = 0$~\cite{Bergman:2010gm}.

By \eqref{7p1}-\eqref{7p3}, the leading DBI \eqref{GenDBI} and WZ \eqref{uuuu} terms are written as
\begin{equation} \label{7pLag}
\begin{split}
  \call_\mathrm{DBI}^{(0)} &= - \sqrt{\Th \k}\sqrt{\left(1+r^4 f z'^2 + r^2 f \psi'^2 - \tA_t'^2 \right)} \,, \\
  \call_{\mathrm{WZ}}^{(0)} &= f_1f_2 r^4 z' - 2 c(r) \tB \tA_t' \,,
\end{split}
\end{equation}
where
\begin{equation} \label{7pk}
\Th = \left(\cos^{4}\psi+\frac{1}{4}f^{2}_{1}\right)
\left(\sin^{4}\psi+\frac{1}{4}f^{2}_{2}\right)\,, \qquad
\kappa=\tilde{B}^{2}+r^{4}\,.
\end{equation}
Note that, in the WZ term, the first term stems from the flux stabilization \eqref{7p1} and
the second term comes from the existence of both density
and magnetic field \eqref{7p2}.
%
%
%

The general OSM expressions \eqref{OSM1}  hold with the following specifications.
\begin{equation}
\begin{split}
& g_{tt}= - r^{2} f(r)\,, \quad  g_{xx} = g_{yy} =r^2 \,,
\quad g_{rr} = \left(\frac{1}{r^{2}f(r)}+r^{2}z^{\prime}(r)^{2} +\psi^{\prime}(r)^{2}\right) \,, \\
& \dd s_{(4)}^2 = \frac{\cos^{4}\psi+\frac{1}{4}f^{2}_{1}}{\cos^{2}\psi}(d\Omega^{(1)}_{2})^2  +\frac{\sin^{4}\psi+\frac{1}{4}f^{2}_{2}}{\sin^{2}\psi}
   (d\Omega^{(2)}_{2})^2\,, \\
&\phi = 0 \,, \qquad  \hJ_t \ra \bar{J}_t=\tilde{J}_{t}+  \frac{c(\psi)}{2}\tilde{B} \,,
\end{split}
\end{equation}
together with \eqref{7pk}.
%
The conductivity for a black hole embedding reads, by the formula with OSM \eqref{sij2D1},
\begin{equation} \label{7pcond1}
\begin{split}
\sigma^{xx}
&=\frac{\mathcal{N}^{\prime}r^{2}_{s}}{\tilde{B}^{2}+r^{4}_{s}}
\sqrt{\bar{J}_t^{2}
+\left(\cos^{4}\psi+\frac{1}{4}f^{2}_{1}\right)
\left(\sin^{4}\psi+\frac{1}{4}f^{2}_{2}\right)(\tilde{B}^{2}+r_{s}^{4})}\,,\\
\sigma^{xy}&= - \mathcal{N}^{\prime}\left(\frac{\tilde{B}\bar{J}(r_{s})}{\tilde{B}^{2}+r^{4}_{s}}
+\frac{c(r_{s})}{2}\right) \,,
\end{split}
\end{equation}
where $\caln' = N_f T_{D7} (4\pi)^2 (2\pi\a')^2$ \eqref{Ns} since $V_{(4)} = (4\pi)^2$. The conductivity for a Minkowski embedding, by \eqref{sHall},
reads
\begin{equation} \label{7pcond2}
  \s^{xx} = 0 \,, \qquad \s^{xy}
  = - \mathcal{N}^{\prime}\frac{c(r_{0})}{2}
  = \caln' \frac{\hJ_t}{\tB} \,,
\end{equation}
which describes a holographic fractional quantum Hall effect.
It is shown that $c(r_0)$ exhibits discrete values originated
from the Dirac quantization of the magnetic flux on the $S^2$ \cite{Bergman:2010gm}. The conductivities \eqref{7pcond1} and \eqref{7pcond2} obtained by OSM method agree to \cite{Bergman:2010gm},
where the real-action method was used~\footnote{A few differences in
coefficients from \cite{Bergman:2010gm} are due to different normalization conventions. $\caln \ra 4 \caln$ and $\bar{J} \ra \frac{\tilde{d}}{4}$ are needed for the same convention as in \cite{Bergman:2010gm}. The factor $4$ difference comes from how to treat
the internal space volume. We include a
whole internal volume $16\pi^2$ to $\caln$, while $4\pi^2$ is
included in $\caln$ in \cite{Bergman:2010gm}.
}.

\paragraph{D2-D8$^\prime$}
Another holographic quantum Hall model at integer filling was constructed in \cite{Jokela:2011eb,Jokela:2011sw}
where the authors considered D2-D8$^\prime$ brane configuration.
\begin{equation}
\label{D2D8p}
\begin{tabular}{ccccccccccc}
& 0 & 1 & 2 & 3 & 4 & 5 & 6 & 7 & 8 & 9 \\
D2 & $\bullet$ & $\bullet$ & $\bullet$ &  & & & & &  \\
D8 & $\bullet$ & $\bullet$ & $\bullet$ & $\bullet$ & $\bullet$ & $\bullet$ & $\bullet$
&  $\bullet$ & $\bullet$ &  \\
\end{tabular}
\end{equation}
Here the D8 brane
wraps $S^{2}\times S^{3}$ inside $S^{6}$ and this configuration
is also nonsupersymmetric and unstable. Thus we have to introduce the following
magnetic field on the internal $S^{2}$ to ensure the stability
\begin{equation} \label{8p1}
2\pi\a'F=\tilde{h}d\Omega_{2}\,,
\end{equation}
where $\dd \Omega_2 \equiv \sin \theta \dd \theta \wedge \dd \phi$
For a detailed argument for this stabilization, we refer to \cite{Jokela:2011eb}.
In addition, we consider the gauge fields
\begin{equation} \label{8p2}
2\pi\a'A=\tilde{A}_{t}\dd t+\tilde{B}x \, \dd y \,,
\end{equation}
representing the charge density and the magnetic field.
With an assumption that the scalar $\psi(=x^9)$ is a
function of only the radial coordinate, $r$,
the pull-back of the metric and the RR 4-from field on the probe D8-brane reads
\begin{equation} \label{8p3}
\begin{split}
ds^{2}_{D8}&=r^{\frac{5}{2}}(-f(r) \dd t^{2}+ \dd x^{2}+ \dd y^{2})+r^{-\frac{5}{2}}
\left(\frac{1}{f}+r^{2}\psi^{\prime2}\right) \dd r^{2} \\
& + r^{-\frac{1}{2}}\sin^{2}\psi \dd \Omega^{2}_{2}+r^{-\frac{1}{2}}\cos^{2}\psi \dd\Omega^{2}_{3}\,, \\
C_5&=c(r) \dd\Omega^{2}_{2}\wedge \dd\Omega^{2}_{3}\,, \quad
c(r) \equiv  c(\psi(r))=\frac{5}{8}\left(\sin\psi-\frac{1}{6}\sin(3\psi)-\frac{1}{10}\sin(5\psi)\right)\,,
\end{split}
\end{equation}
where $f(r)=1-r^{5}_{H}/r^{5}$.
Note that the dilaton is nontrivial $e^{-\phi}=r^{-5/4}$.

By \eqref{8p1}-\eqref{8p3}, the DBI \eqref{GenDBI} and WZ \eqref{uuuu} terms are written as
\begin{equation}
\begin{split}
\call_{\mathrm{DBI}}^{(0)} &=  -\sqrt{\Th \k}\sqrt{1+r^2f\psi'^2 - \tA'^2}        \,, \\
\call_{\mathrm{WZ}}^{(0)}  &= -  c(r)\tilde{A}^{\prime}_{t}\tilde{B}\,,
\end{split}
\end{equation}
where
\begin{equation} \label{8pk}
\Theta = \left(r^{-1}\sin^{4}\psi+\tilde{h}^{2}\right)r^{-\frac{3}{2}}\cos^{6}\psi\,, \qquad
\kappa=\tilde{B}^{2}+r^{5}\,.
\end{equation}

The general OSM expressions \eqref{OSM1} hold with the following specifications.
\begin{equation}
\begin{split}
& g_{tt}= - r^{5/2} f(r)\,, \quad  g_{xx} = g_{yy} =r^{5/2} \,,
\quad g_{rr} = r^{-\frac{5}{2}}
\left(\frac{1}{f}+r^{2}\psi^{\prime2}\right) \,, \\
& \dd s_{(5)}^2 = \frac{r^{-1}\sin^{4}\psi+\tilde{h}^{2}}{r^{-\frac{1}{2}}\sin^{2}\psi} \dd \Omega_2^2  + r^{-\frac{1}{2}}\cos^{2}\psi \dd\Omega^{2}_{3}    \dd\Omega^{2}_{3}\,, \\
&e^{-\phi}=r^{-5/4}\,, \qquad  \hJ_t \ra \bar{J}_t=\tilde{J}_{t}+  c(\psi)\tilde{B} \,.
\end{split}
\end{equation}
%
%
%
%
%
The conductivity for a black hole embedding reads, by the formula with OSM \eqref{sij2D},
\begin{equation} \label{8pcond1}
\begin{split}
\sigma^{xx}&= \caln' \frac{r^{5/2}_{s}}{\tilde{B}^{2}+r^{5}_{s}}
\sqrt{\bar{J}_t^{2}+\cos^{6}\psi_{s}(\tilde{h}^{2}r_{s}+\sin^{4}r_{s})(r^{5}_{s}+\tilde{B}^{2})}\,,\\
\sigma^{xy}&= \caln' \left(\frac{\bar{J}_t\tilde{B}}{r^{5}_{s}+\tilde{B}^{2}}+c(r_{s})\right)\,,
\end{split}
\end{equation}
where $\caln' = N_f T_{D8} 8\pi^3 (2\pi\a')^2  $ \eqref{Ns} since $V_{(5)} = 8\pi^3$. The conductivity for a Minkowski embedding, by \eqref{sHall},
yields
\begin{equation} \label{8pcond2}
  \s^{xx} = 0 \,, \qquad \s^{xy}
  = - \mathcal{N}^{\prime} c(r_{0})
  = \caln' \frac{\hJ_t}{\tB}  = \frac{N}{2\pi} \,,
\end{equation}
which describes a holographic integer quantum Hall effect
($N$  is the integer filling fraction, $\caln' = 3 N/4\pi$ \cite{Jokela:2011eb}).
The conductivities \eqref{8pcond1} and \eqref{8pcond2} obtained by OSM method agree to \cite{Jokela:2011eb},
where the real-action method was used.

\subsection{Light-cone AdS black hole: Strange metal}
\label{LigthCone}

AdS space in the light-cone frame (ALCF) is proposed as a physical system belonging to the universality class of the normal state of unconventional superconductor, showing its universal
conductivity properties: linear temperature dependent Ohmic resistivity and quadratic temperature dependent inverse Hall angle \cite{Kim:2010zq}.

The metric of ALCF \eqref{ALCFmetric} has an off-diagonal component
as in \eqref{eq.nondiagmet}, so it serves as a good example to apply
our general formalism for the conductivity in terms of OSM \eqref{LCcond11}.
%
%
%
%
%
%
The ALCF metric is obtained from AdS$_5 \times S^5$ metric by the transformation $x^{+}=b(t+x)$, $x^{-}=\frac{1}{2b}(t-x)$, which yields
\begin{equation} \label{ALCFmetric}
\begin{split}
\dd s^{2}=&\, g_{++}\dd x^{+2}+2g_{+-}\dd x^{+}\dd x^{-}+g_{--}\dd x^{-2}+g_{yy}\dd y^{2}+g_{zz}\dd z^{2}+g_{rr}\dd r^{2} \\
&+ R^2 \cos^2\theta \dd \Omega_3^2 + R^2 \sin^2\theta \dd \phi^2\,,
\end{split}
\end{equation}
with
\begin{align} \label{met11}
g_{++}&=\frac{(1-f(r))r^{2}}{4b^{2}R^{2}}\,,& g_{+-}&=-\frac{1+f(r)r^{2}}{2R^{2}}\,,& g_{--}&=\frac{(1-f(r))b^{2}r^{2}}{R^{2}}\,,& \nn \\
g_{yy}&=g_{zz}=\frac{r^{2}}{R^{2}}\,,&  g_{rr}&=\frac{R^{2}}{r^{2}f(r)}\,,&   f(r)&=1-\frac{r^{4}_{H}}{r^{4}}\,,&
\end{align}
where $R$ is AdS$_5$ radius and $b$ is the parameter related to the rapidity and $r_H$ is the horizon position.
The worldvolume $U(1)$ gauge fields for density and magnetic field are~\cite{Kim:2010zq}
\begin{equation}
2\pi\alpha'A  = \th_{+}(r)  \dd x^+
+ \th_{-}(r)  \dd x^- + \tB_b \, y \ud z ,
\end{equation}
where, again, variables with tilde include a $2\pi\alpha'$ factor.

We introduce $N_f$ D7 branes and consider a non-trivial embedding scalar $\theta(r)$ with the other scalar $\phi=0$. The DBI action, then, reads
\begin{eqnarray}
S_{D7} = -N_{f}T_{D7}\int \dd^{8}\xi\sqrt{-{\rm det}(g_{D7}+2\pi\alpha^{\prime}F)} = \caln \int \dd^{5}\xi \, \mathcal{L} \ ,
\end{eqnarray}
where $\mathcal{N}\equiv2\pi^{2}N_{f}T_{D7}$ and
\begin{equation}
\mathcal{L} = -  \sqrt{( \tB^2 + G_{yy} g_{zz}) g_{\Omega\Omega}^{3}}
 \sqrt{-\mathfrak{g} g_{rr}^{D7} - g_{--}g_{yy}\th^{\prime2}_{+}
-g_{++}g_{yy}\th^{\prime2}_{-}
+2 g_{+-}g_{yy}\th^{\prime}_{+}\th^{\prime}_{-} } \,,
\end{equation}
with
\begin{equation}
   g_{rr}^{D7}=g_{rr}+R^{2}\theta^{\prime}(r)\,, \qquad g_{\Omega\Omega}=R^{2}\cos^{2}\theta \,, \qquad \mathfrak{g} = g_{+-}^2 - g_{++} g_{--} \,.
\end{equation}
There are two conserved currents conjugate to two cyclic coordinates
$\th_+$ and $\th_-$:
\begin{equation} \label{conc}
\begin{split}
\hJ_{+}&=\frac{\partial\mathcal{L}}{ \partial \th^{\prime}_{+}}
=\frac{g_{zz}g_{\Omega\Omega}^{3}}{\mathcal{L}}
\left(g_{--}g_{yy}\th^{\prime}_{+}-g_{+-}g_{yy}\th^{\prime}_{-}\right)\,,\\
\hJ_{-}&=\frac{\partial\mathcal{L}}{ \partial \th^{\prime}_{-}}
=\frac{g_{zz}g_{\Omega\Omega}^3}{\mathcal{L}}
\left(g_{++}g_{yy} \th^{\prime}_{-}  - g_{+-}g_{yy}\th^{\prime}_{+}\right)\,.
\end{split}
\end{equation}

The OSM \eqref{eq.sab} reads
\begin{equation}
\dd s^{2}=s_{++}\dd x^{+2}+2s_{+-}\dd x^{+}\dd x^{-}+s_{--}\dd x^{-2}+s_{yy}\dd y^{2}+s_{zz}\dd z^{2}+s_{rr}\dd r^{2},
\end{equation}
where
\begin{equation} \label{BomOSM}
\begin{split}
s_{++} &= g_{++} + \frac{\fg_1^2}{\chi}\,, \quad
s_{+-}  = g_{+-} + \frac{\fg_1 \fg_2}{\chi}\,, \quad
s_{--}  = g_{--} + \frac{\fg_2^2}{\chi} \,, \\
s_{yy} &= g_{yy} + \frac{\tB}{g_{zz}}\,, \,\,\quad
s_{zz}  = g_{zz} + \frac{\tB}{g_{yy}}\,, \ \,\quad \quad
s_{rr}  =  \frac{\xi_1 g_{\Omega\Omega} g_{rr}^{D7}}{\chi} \,,
\end{split}
\end{equation}
with
\begin{equation} \label{xi1}
\begin{split}
  \chi &= \frac{\xi_1 g_{--} g_{\Omega\Omega}^3 - \fg_2^2 +
  \mathfrak{g} J_{+}^2 }{ g_{--} } \,, \quad
  \xi_1 = \mathfrak{g} (\tB^2 + g_{yy} g_{zz} ) \,, \\
  \mathfrak{g} &= g_{+-}^2 - g_{--} g_{++} \,, \quad
  \fg_1 = g_{+-}J_{-} + g_{++}J_{+} \,, \quad
  \fg_2 = g_{--}J_{-} + g_{+-}J_{+} \,.
\end{split}
\end{equation}
Note that the background gauge field information is geometrized in
$\chi,\chi_1, \fg_1, \fg_2$
and the effective coupling
\begin{equation}
  g_5 = \sqrt{\frac{g_{yy}g_{zz}\chi \fg}{\xi_1^2 g_{\Omega\Omega}^3}} \,.
\end{equation}

To compute a nonlinear conductivity we need to introduce an electric field, which generates a singular shell. Let us consider an electric field along
$y$ direction, $F_{y+} = E_b$, and find the singular shell position by \eqref{xx}\footnote{If we started with a D5 probe brane, we would not
have the singular shell at all.}:
\begin{equation} \label{rsBom}
  \xi(r_s) = \det \g_{\m\n} = \tE_b^2 g_{--}(r_s)g_{zz}(r_s) - \xi_1(r_s) = 0 \,,
\end{equation}
which yields
\begin{equation} \label{sing11}
   r_s = \left(2 b^2 \tE_b^2 R^4 \left(t^4 -\calb^2 + \sqrt{t^4 + (\calb^2 + t^4 )^2} \right) \right)^{1/4} \,,
\end{equation}
where
\begin{equation}
  t=  \frac{\pi R b T}{\sqrt{2b\tE_b}}\,, \qquad
  \calb  = \frac{\tB}{2b \tE_b} \,.
\end{equation}

Plugging OSM \eqref{BomOSM} and \eqref{sing11} into \eqref{LCcond11} we have
\begin{equation}\label{Bom11}
\begin{split}
  \sigma^{yy} &= \sigma_0 \frac{\sqrt{\calf_- J^2 + t^4 \sqrt{\calf_-} \calf_+}}{\calf_+} \,,   \\
  \sigma^{yz} &= \bar{\sigma}_0 \frac{\calb}{\calf_+} \,,
\end{split}
\end{equation}
where
\begin{align}
  \calf_{\pm} & = \frac{\sqrt{(\calb^2 + t^4)^2 + t^4} \pm \calb^2 + t^4}{2} \,,&
 J & = \frac{\hJ^+}{ R^3 b \cos^3 \theta(r_s) (2 b \tE)^{3/2} }\,,& \\
    \s_0 &  = \caln' R^3 \sqrt{2 b^3 \cos^6 \theta(r_s) \tE_b} \,,&
    \bar{\s}_0 & = \caln'\frac{\hJ_+}{b\tE_b}  \,.&
\end{align}
This agrees to \cite{Kim:2010zq} and \eqref{RA_Bomsoo} obtained by real-action method. At $B=0$
\begin{equation} \label{AAA}
  \calf_{+} = \calf_{-} = t^2  \cala \,, \qquad \cala = \frac{t^2 + \sqrt{1+t^4}}{2} \,,
\end{equation}
and the Ohmic conductivity is simplified as
\begin{equation} \label{Bom22}
\s^{yy} = \sigma_0 \sqrt{\frac{J^2}{t^2 A(t)} + \frac{t^3}{\sqrt{A(t)}}} \,.
\end{equation}

The Ohmic conductivity, $\s^{yy}$, consists of two parts: a finite charge density contribution (the term with $J$) and a thermally created charge pair contribution. Note that the embedding information ($\theta(r_s)$) related to the mass of the charge carrier is contained in $\s_0$ and $J$.
However, since this contribution cancels out in $\s_0 J$, the conductivity
due to a finite charge carrier does not depend on the embedding or
the mass of the charge carrier. The conductivity due to pair creation has
a dependence on the mass of the charge carrier through $\s_0$.
The large mass corresponds to the embedding $\theta_s \sim \pi/2$, so the 
pair produced conductivity by the large mass charge carriers will be Boltzmann suppressed.

%
At $B=0$, in the regime $t \ll J^{1/3}$ and $1 \ll J $, the conductivity \eqref{Bom22} is dominated by the first term
\begin{equation} \label{St1}
  \s^{yy} \sim  \frac{\hJ^+}{t\sqrt{t^2 + \sqrt{1+t^4}}}  \sim
  \begin{cases}
   \hJ^+ /t  &  t \ll 1 \\
   \hJ^+ /t^2 & t \gg 1
  \end{cases}\,,
\end{equation}
where $t$ can be tuned by changing $b$ at fixed $\tE_b$ and $RT$.
At $t \ll 1$ we obtained the resistivity linear in temperature. Interpreting $b$ as a doping parameter we see a typical cross over behavior of the strange metal. At $B \ne 0$, in the regime, $t\ll\sqrt{\calb}$, $t \ll \frac{J}{\calb}$, and $\calb \gg 1$ the conductivity \eqref{Bom11} is approximated as
\begin{equation} \label{St2}
  \s^{yy} \sim \frac{\hJ^+ t^2}{\calb^2} \,, \qquad
  \s^{yz} \sim \frac{\hJ^+}{\calb} \,, \quad \lra \quad
  \frac{\s^{yy}}{\s^{yz}} \sim \frac{t^2}{\calb} \,,
\end{equation}
where the Ohmic conductivity is dominated by the first term.
The temperature dependence ($\sim t^2$) of the inverse Hall angle
is the typical property of the strange metal.
Note that, if $\calb \gg J$ ($t \ll 1$), the ohmic conductivity is $1/T$,
and, if $J \gg \calb$ ($t \ll \infty$), it is possible to cross over to $1/T^2$ \eqref{St1}.
Note that the ALCF model may show strange metallic behaviors \eqref{eq.1} for certain parameter regime \cite{Kim:2010zq}:
\begin{equation} 
  \s^{xx} \sim \frac{1}{T} \,, \quad
  \s^{xy} \sim \frac{1}{T^3} \,, \quad
  \cot \theta_H \sim \frac{\s^{xx}}{\s^{xy}} \sim T^2 \,. \nn
\end{equation}
%

\section{Properties of holographic DC conductivity: towards strange metal}

Since the ALCF model shows strange metallic behaviors \eqref{eq.1} in a simple framework,
it is worth while to analyze what makes ALCF to be different from other models, for example, Lifshitz geometry, where there is some ``no-go'' argument
for strange metal \cite{Hartnoll:2009ns}, which we will review in the following subsection. i.e. What property of ALCF makes ``yes-go''? This analysis will be useful also for a model-building towards
holographic strange metal.

For this purpose, let us start by expressing our general conductivity formulae obtained by OSM method, in terms of the original metric, conserved charge and background electromagnetic field. In 2+1 dimensions
\begin{equation} \label{gg1}
\begin{split}
\sigma^{ii}&= \left.\mathcal{N}' \frac{g_{xx} g_{yy}}{g_{ii}}
\frac{\sqrt{e^{-2\phi} \Th (g_{xx}g_{yy}+\tB^2) + (\hJ_{t} + C_{q-3} \tB )^2     }}{g_{xx}g_{yy}+\tB^2}\right|_{r \ra r_s} \,, \\
\sigma^{ij}&= \mathcal{N}' \epsilon^{ij} \left. \frac{\tB \hJ_{t} -   C_{q-3} }{g_{xx}g_{yy}+\tB^2}\right|_{r \ra r_s}\,,
\end{split}
\end{equation}
which agrees to \eqref{A1} obtained by real-action method, and for a diagonal metric in 3+1 dimensions
\begin{equation} \label{gg2}
\begin{split}
 \s^{ii} &= \left. \caln' \left(\tB_i^2+ \frac{g_{xx}g_{yy}g_{zz}}{g_{ii}}\right)
\frac{\sqrt{\hJ_t^2 + e^{-2\phi} \Th \k } }{\k }\right|_{r \ra r_s} \,,  \\
 \s^{ij} &= \left. \caln' \tB_i \tB_j
\frac{\sqrt{\hJ_t^2 +  e^{-2\phi} \Th \k } }{\k } - \caln' \e^{ijk} \frac{\tB_k g_{kk} \hJ_t}{\k} \right|_{r \ra r_s} \,,
\end{split}
\end{equation}
where $\k = g_{xx}g_{yy}g_{zz} +  \sum_{i=1}^3 g_{ii}^2 B_i^2 $ \eqref{ka3}. This agrees to (\ref{genc})\footnote{Indeed \eqref{gg2} is more general
than \eqref{genc}. For a direct comparison with \eqref{genc} we
should set $g_{xx} = g_{yy} = g_{zz}$ and $\tB_y$=0.} obtained by
the real-action method.

\subsection{AdS light-cone frame}

We start with a review on how to read off temperature
dependence from the metric~\cite{Hartnoll:2009ns,Pal:2010sx}.
At large density, \eqref{gg1} and \eqref{gg2} are dominated by
\begin{eqnarray} \label{asymp1}
  \s^{xx} \sim \frac{\hJ^t}{g_{xx}(r_s)}   \,, \qquad \frac{\s^{xx}}{\s^{xy}} \sim \frac{g_{yy}(r_s)}{ \tB} \sim \frac{g_{xx}(r_s)}{ \tB} \,,
\end{eqnarray}
where we consider the case $g_{xx} = g_{yy}$.
If the theory is invariant under the form
\begin{equation}
  t \ra \l^z t \,, \qquad \vec{x} \ra \l \vec{x} \,,
\end{equation}
we assign a momentum dimension to $t$ and $x$ as $[t]=-z\,, [\vec{x}]=-1$.
Then $[\s] = d-2$, $[\hJ^t] = d$, $[\tB] = 2$, and
$[T] = z$ with $\hbar = k_B = e =1$.  Consequently $[g_{xx}] = 2 \sim T^{2/z}$.
We can arrive at only one strange metal property: either $\s^{xx} \sim 1/T$ ($z=2$) or $\s^{xx}/\s^{xy} \sim T^2$ ($z=1$). However, since $\s^{xx}/\s^{xy} \sim (\s^{xx})^{-1}$ for all $z$, we cannot obtain both\footnote{
In \cite{Pal:2010sx}, with an assumption of anisotropic scaling $g_{xx} \ne g_{yy}$, $t \ra \l t , x \ra \sqrt{\l}x, y \ra \l y$, it was claimed that the strange metal properties could be modeled.
Anisotropy of the metric may mean anisotropy of the medium.}. Let us call this ``no-go'' argument.

However, this argument is based on the diagonal background metric, while ALCF metric has off-diagonal components. So we may suspect that
the strange metal properties may be related to off-diagonal metrics.
To see this, we rewrite the conductivities \eqref{LCcond} in terms of a general metric.
\begin{equation}
\begin{split} \label{ggg1}
  \sigma^{yy} &= \caln' g_{zz}\left. \frac{\sqrt{ \frac{g_{--}}{g_{zz}} g_{\Omega\Omega}^3(g_{yy}g_{zz}+\tB^2)
  +  \hJ_+^2}}{g_{yy}g_{zz}+\tB^2} \right|_{r\ra r_s} \,,  \\
    \sigma^{yz} &= \caln' \left. \frac{\tB \hJ_+  }{g_{yy}g_{zz}+\tB^2}\right|_{r\ra r_s} \,,
\end{split}
\end{equation}
which agrees to \eqref{RA_Bomsoo} obtained by the real-action method.
At large density, \eqref{ggg1} are dominated by
\begin{eqnarray} \label{ALCF1}
  \s^{yy} \sim \frac{\hJ^t}{g_{yy}(r_s)}   \,, \qquad \frac{\s^{yy}}{\s^{yz}} \sim  \frac{g_{yy}(r_s)}{ \tB} \,,
\end{eqnarray}
where $g_{yy} = g_{zz}$. Interestingly, this has the same form as \eqref{asymp1} even though we started with metric with non-zero off-diagonal components\footnote{Where is, then, the effect of off-diagonal metric? It is encoded in
the singular shell position, $r_s$ \eqref{rsBom}. See $\xi_1$ and $\fg$
in \eqref{xi1}. Note that $g_{yy}$ has all information on scaling and
it is simply $\sim r_s^2$. }.
So ALCF's strange metal behavior looks contradictory to ``no-go'' argument.
However, there is no contradiction, since ALCF's computation involves more scales than temperature, while ``no-go''argument assumes only one scale, temperature.

From \eqref{met11} and \eqref{sing11} we have
\begin{align}
   g_{yy}(r_s) \sim r_s^2 &\sim  \sqrt{t^4 -\calb^2 + \sqrt{t^4 + (\calb^2 + t^4 )^2} }  \label{non11} \\
   & \sim \begin{cases} \label{ttt}
   t  &  \ \calb = 0, \ t \ll 1 \\
   t^2 & \ \calb = 0, \ t \gg 1 \\
   t^2 & \ \sqrt{\calb} \gg t , \ \calb \gg 1
  \end{cases}\,,
\end{align}
which explicitly shows what is going on and how ``yes-go'' is possible.
It is important to note that a new controlling parameter $b$, to be
identified as a doping parameter, made this possible, since this new
dimensionful parameter enabled the electric field and the magnetic field to
enter the game. For example, $t \ll 1$ does not necessarily mean
$T \ll \sqrt{\tE_b}$. It may be the case, $T \gg \sqrt{\tE_b}$ but with a small enough $b$. For ``no-go'' argument, it was always assumed
that $T \gg \sqrt{\tE_b}$ and $T \gg \sqrt{\tB}$ so the {\it non-linear}
conductivity formula such as \eqref{non11} cannot play a role.
Having $\tE = \tB = 0$, the conductivity playing a role, is essentially  
a {\it linear} one. 
However in the ALCF's case, thanks
to $b$, the {\it non-linear} conductivity formula is still important
even in the case $T \gg \sqrt{\tE_b}$ and $T \gg \sqrt{\tB}$.

To compare the differences from other cases, let us consider
D3/D7(D5) case, where
\begin{equation} \label{gyy11}
  g_{yy}(r_s) \sim  \sqrt{T^4 -\tB^2 + \tE^2 + \sqrt{-4\tB^2 \tE^2 + (T^4 + \tB^2 + \tE^2)^2}} \, .
\end{equation}
In the limit for ``no-go'' argument (small $\tE$ and $\tB$) $g_{yy}(r_s) \sim T^2$. However, in order to see an effect of additional parameter such as $b$ in the ALCF model, we consider the other limit (large $\tE$ and/or $\tB$) so that another scale $\tE$ or $\tB$ can play a role. This is only a theoretical consideration to understand a mathematical structure, since
we are eventually interested in a small electromagetic regime.  The question we want to ask is ``is it possible to have two scalings as in  \eqref{ttt} by tuning $\tE$ or $\tB$?''.   It turns out that $g_{yy}(r_s) \sim T^0$ or $\sim T^2$ and does not work.
For the Lifshitz case with a dynamical exponent $z=2$, we considered two black hole factors $1-r_H^2/r^2$ and $1-r_H^4/r^4$. In these cases, $g_{yy}(r_s)$ are
very complicated functions, but in the extreme limit $\tE \gg 1$ and/or $\tB \gg 1$, the structure is similar to \eqref{gyy11} and does not
work either.
Thus, simply introducing an additional parameter is not sufficient to have a structure like \eqref{ttt}.

\subsection{Charged dilatonic black holes}

There exist a large class of charged dilatonic black holes and it was argued in~\cite{Hartnoll:2009ns} that
a non-trivial dilaton might lead to a more promising holographic model building for strange metals.  Several
attempts were carried out in e.g.~\cite{Charmousis:2010zz,Lee:2010qs,Lee:2010ii,Lee:2010uy}.  As our final example for the OSM method, we consider four dimensional charged dilatonic solution in~\cite{Gubser:2009qt}:

\begin{equation} \label{gubser1}
\mathcal{L}=\frac{1}{2\kappa^{2}}\left[R-\frac{1}{4}e^{\alpha}F_{\mu\nu}F^{\mu\nu}-\frac{3}{2}
\partial_{\mu}\alpha\partial^{\mu}\alpha+\frac{6}{L^{2}}\cosh\alpha\right]\,,
\end{equation}
of which classical solution is
\begin{equation}
\begin{split}
\dd s^{2}& = e^{2A}\left(-h\dd t^{2}+\dd x^{2}+\dd y^{2}\right)
+\frac{e^{2B}}{h}\dd r^{2}\,,\\
A &=\log\frac{r}{L}+\frac{3}{4}\log\left(1+\frac{Q}{r}\right)\,, \quad
  B=-A\,, \quad h=1-\frac{\mu L^{2}}{(Q+r)^{3}}\,,\\
\alpha & =\frac{1}{2}\log\left(1+\frac{Q}{r}\right)\,, \quad
A_{t}=\frac{\sqrt{3Q\mu}}{Q+r}-\frac{\sqrt{3Q}\mu^{\frac{1}{6}}}{L^{\frac{2}{3}}}\,,
\end{split}
\end{equation}
The temperature is
\begin{equation}
T=\frac{1}{4\pi}\frac{1}{\sqrt{-g_{tt}g_{rr}}}\frac{d}{dr}g_{tt}|_{r=r_{H}}
=\frac{3\mu^{1/6}r_{H}^{1/2}}{4\pi L^{5/3}}\,,
\end{equation}
where the horizon, $r_H$, is at $\mu^{1/3}L^{2/3}-Q $ such that $h(r_{H})=0$.  As investigated in~\cite{Gubser:2000mm}, such a solution is the dimensional reduction of the 4-charge
$AdS_{4}\times S^{7}$ black hole solution in eleven dimensional supergravity down to four dimensions with
equal charges.  For simplicity we neglect the details of the embedding into string theory, even though it is possible, and just assume a dilaton field $\phi$ and some internal space.

The conductivities, \eqref{gg1}, omitting the CS terms, are
\begin{equation}
\begin{split}
\sigma^{xx}& \left. =\frac{\mathcal{N}^{\prime}e^{2\phi}g_{xx}
\sqrt{e^{-2\phi}g^{n}_{\Omega\Omega}(e^{4\phi}g_{xx}g_{yy}+\tilde{B}^{2})+\hat{J}^{2}_{t}}}
{e^{4\phi}g_{xx}g_{yy}+\tilde{B}^{2}}\right|_{r\rightarrow r_{H}}\,, \\
\sigma^{xy}& \left. =\frac{\mathcal{N}^{\prime}\tilde{B}\hat{J}_{t}}{e^{4\phi}g_{xx}g_{yy}+\tilde{B}^{2}}
\right|_{r\rightarrow r_{H}}\,,
\end{split}
\end{equation}
where we substituted $g_{\m\n} \ra e^{2\phi} g_{\m\n}$, since \eqref{gubser1} are written in Einstein frame.

At large density and small magnetic field limit, the conductivities read
\begin{equation}
  \s^{xx} \sim \frac{\hJ^t}{e^{2\phi}g_{yy}} \,, \qquad
  \frac{\s^{xy}}{\s^{xx}} \sim \frac{e^{2\phi} g_{xx}}{\tB} \,,
\end{equation}
so a dilaton field does not help to avoid the ``no-go'' argument, but it affects
the conductivity individually \cite{Hartnoll:2009ns,Pal:2010sx}.
By assuming $\phi \sim \a$ and $\mu \gg Q$, we have
\begin{equation}
  \s^{xx} \sim \frac{1}{T^2} \,, \qquad \frac{\s^{xy}}{\s^{xx}} \sim T^2\,,
\end{equation}
while by assuming $\phi \sim -1/6\log\mu $
\begin{equation}
  \s^{xx} \sim \frac{1}{T} \,, \qquad \frac{\s^{xy}}{\s^{xx}} \sim T \,.
\end{equation}
In the former, only Hall angle matches the strange metal property, while
the latter, only Ohmic conductivity does.

\section{Open string metric beyond the DC conductivity and discussions}
\label{sec.7}

We studied the holographic DC conductivities of various systems using
the OSM method. There are two main points: formalism and applications to strange metal.

First, we proposed a new method to compute the DC conductivity based on OSM. We showed that all results obtained by the OSM method agreed to the results obtained by the real-action method.
Therefore, the OSM method is equivalent to the real-action method as
far as the final conductivity formula is concerned.
However, it gives us a new conceptual insight.
For example, the {\it field theoretic} real-action condition can be interpreted as the {\it geometric} regularity condition of OSM (section \ref{3.2}).
It also yields a technical simplification, since most computations are
simply matrix operations (the last paragraph of section \ref{sec.4}).
Furthermore, as we will discuss below, OSM can be used to study other transport coefficients and effective temperature induced by the effective world volume horizon, contrary to the real-action method.

Second, we analyzed the conductivity formulae written in terms of general metric, density and electromagnetic fields in order to see
how we can model holographic strange metal properties in general.
(These general forms have been obtained by the OSM method but
also confirmed by the real-action method in the appendix.).
We found that the formula \eqref{ALCF1}
\begin{eqnarray} \label{bbbb}
  \s^{yy} \sim \frac{\hJ^t}{g_{yy}(r_s)}   \,, \qquad \frac{\s^{yy}}{\s^{yz}} \sim  \frac{g_{yy}(r_s)}{ \tB} \,, 
\end{eqnarray}
holds both for the Lifshitz background and the Light-cone AdS background.
So it looks that we are doomed to have $\s^{xx}/\s^{xy} \sim (\s^{xx})^{-1}$, which mimics the Drude result rather than a strange metal. 
Indeed it was shown to be the Drude result for the Lifshitz background \cite{Hartnoll:2009ns}. 
However, having the other parameter $b$, to be identified to the doping parameter, the Light-cone AdS space can manage to have $\s^{xx}/\s^{xy} \ne  (\s^{xx})^{-1}$ and show the strange metal property \eqref{eq.1}.

In this paper, we focused on the DC conductivity as an application
of the OSM.
It was particularly simple because the DC conductivity is defined
in the limit of zero momentum, $k^\m \ra 0$, where the $r$-flow equation of the conductivity becomes trivial and the linear response at the horizon is enough to determine the linear response of the boundary field theory.
However, at finite frequency and momentum (the next order in the derivative expansion) the flow equation should be integrated up to
the boundary and all bulk information becomes relevant. Typical examples are AC conductivity, viscosities, the charge diffusion constant and charge susceptibility.
Also in these cases, the OSM will be a useful starting point.

As an immediate application,
let us comment on the diffusion constant  ($D$)  and charge susceptibility  ($\Xi$).
It was shown, from the membrane paradigm method \cite{Iqbal:2008by}, that the diffusion constant and charge susceptibility could be expressed as integrations of the flow equation of the longitudinal conductivity and Maxwell equations respectively:
\begin{eqnarray} \label{DXi}
D =\sigma\int^{\infty}_{r_{0}}dr\frac{g_{tt}g_{rr}}{\sqrt{-g}}g^{2}_{d+1}\,, \qquad
\Xi =  \left[\int^{\infty}_{r_{0}}dr\frac{g_{tt}g_{rr}}{\sqrt{-g}}g^{2}_{d+1} \right]^{-1} \,,
\end{eqnarray}
which obeyed the Einstein relation, $\Xi D = \sigma$.
This formula is based on two assumptions (1) the metric $g_{\mu\nu}$ is
diagonal and (2) The gauge field dynamics is determined only by the Maxwell terms. Thus, if the OSM satisfies these assumptions, we simply use \eqref{DXi} by replacing $g_{\m\n} \ra s_{\m\n}$.
Two examples are the case with $B \ne 0, d = 0$ and $d \ne 0, B =0$.
See, for example, \eqref{OSM1}, where the $s_{\m\n}$ is diagonal
and $Q=0$. So we can easily compute the diffusion constant and the
charge susceptibility at finite $B$ or $d$ using the OSM
and it proves that the Einstein relation holds even at finite $B$ or $d$.
This is a non-trivial new result, even though it looks straightforward
from OSM point of view. Furthermore, this OSM argument also easily shows
the universality of the Einstein relation with certain
finite background field satisfying the above mentioned assumptions.
When both $B$ and $d$ are nonzero, \eqref{DXi} will be modified
a little bit from the contribution of $Q$ in \eqref{OSM1}, but
we suspect that the Einstein relation should still hold.

However, there is a subtlety related to the scalar modes of the probe brane system.
As discussed in the footnote \ref{mmm}, the probe brane embedded in higher dimension has scalar modes in addition to gauge field vector modes so
there may be couplings between scalar and gauge fields.
Since this coupling is model dependent, it should be
dealt with case by case. For example, in the D3/D7 model,
there is a coupling between the longitudinal gauge field and a scalar
field, which violate the assumption (2).
Therefore, the diffusion constant calculation may be affected by a scalar mode. This issue was addressed in \cite{Mas:2008qs} for the D3/D7 at
finite density. It has been shown that there is a decoupling
at zero quark mass case so \eqref{DXi} can be applied safely \cite{Kim:2008bv, Mas:2008qs}. Therefore we can easily see the Einstein relation holds, which confirms \cite{Mas:2008qs} from OSM point of view. 
At finite quark mass, it was argued that the Einstein relation still holds based on  a numerical analysis \cite{Mas:2008qs}.

Another interesting aspect of the OSM is that it defines an effective
temperature by the surface gravity at the singular shell, which is
nothing but the effective world volume black hole horizon.
In this paper, we did not include
the electric field and currents in the geometry represented by OSM.
The electric field was introduced
only to determine the singular shell position and the
currents were the outputs of our method.
However, if we are interested in the effective geometry back-reacted by the
electric field and currents we have to include them in the OSM as has been done for D3/D7 system in \cite{Kim:2011qh}. It will be interesting to do this analysis in various cases including the Lifshitz geometry.

There is another way of introducing an effective horizon and temperature.
A moving brane or strings develops an effective event or apparent horizon on its world volume \cite{Das:2010yw,Janiszewski:2011ue}.
For example, the effective temperature of the rotating D7 brane
was studied in \cite{Das:2010yw}. Indeed a constant brane motion
is T-dual to the constant electric field, and they both develop a world volume horizon. If we turn on the electric field in
the rotating D7 brane with or without density and/or magnetic field,
the effective horizon and temperature given by the rotation will be
modified by the effect of the electromagnetic field. This will yield richer and interesting thermodynamics of flavors in a Large $N$ gauge theory.

\acknowledgments

\noindent
We would like to thank Nick Evans, Jonathan Shock, Javier Tarrio, Andy O'Bannon, Bom-Soo Kim, Jos\'{e} P. S. Lemos, Matthew Lippert, 
Jian-Huang She, and Sung-Sik Lee for helpful comments and discussions.
K.K. is grateful for University of Southampton Scholarships.
D.W.P. acknowledges an FCT (Portuguese Science Foundation) grant.
This work was also funded by FCT through project PTDC/FIS/098962/2008.

\appendix

\section{DC conductivity by Karch-O'bannon's real-action method}

In this appendix we briefly show how to compute the DC conductivity
by Karch-O'bannon's real-action method \cite{Karch:2007pd}.
All results obtained here agree to the results by the OSM method.

\subsection{General dimension}

The DBI action of $N_f$ probe Dq-branes ($q=d+1+n$) reads
\begin{eqnarray}
S_{Dq}=-N_{f}T_{Dq}\int \dd^{q+1}\xi \, e^{-\phi}\sqrt{-{\rm det}(g_{mn}+(2\pi\alpha^{\prime})F_{mn})} \,. \label{ADBI}
\end{eqnarray}
With the induced metric
\begin{equation}
\dd s^{2}=g_{tt}\dd t^{2}+g_{rr}\dd r^{2}+g_{xx}\sum\limits^{d}_{i=1}\dd x^{2}_{i}+g_{\Omega\Omega}d\Omega^{2}_{n}\,,
\end{equation}
and the gauge field\footnote{ As in the main text,
the gauge fields with tilde include $2\pi\alpha'$ factor.
For example, $\tA_t = 2\pi\alpha' A_t$. }
\begin{equation}
\label{u1g}
2\pi\a'A=\tilde{A}_{t}(r)\dd t+(-\tilde{E}_{x}t+\tilde{A}_{x}(r))\dd x
+(\tilde{B}_{z}x+\tilde{A}_{y}(r))\dd y+(\tilde{B}_{x}y + \tilde{A}_{z}(r))\dd z\,,
\end{equation}
the DBI action \eqref{ADBI} is rewritten as
\begin{align}
S_{Dq}
&= -\mathcal{N}\int \dd^{d+2}\xi \, \call \\
&= -\mathcal{N}\int \dd^{d+2}\xi \, e^{-\phi}g_{\Omega\Omega}^{\frac{n}{2}}\sqrt{-{\rm det}\gamma_{mn}}\,,  \label{detgmn}\\
&=-\mathcal{N}\int \dd^{d+2}\xi \, e^{-\phi}g_{xx}^{\frac{d-3}{2}}g_{\Omega\Omega}^{\frac{n}{2}}
\sqrt{-g_{tt}g_{rr}g_{xx}^{3}-g_{xx}\cala_{2}-\cala_{4}}\,,
\end{align}
where $m,n$ are indices only for $d+2$ space excluding internal sphere and
\begin{equation}
\begin{split}
\mathcal{N}&=N_{f}T_{Dq} \int \Omega_{n}\,, \qquad
\gamma_{mn}=g_{mn}+2\pi\alpha^{\prime}F_{mn}\,, \\
\cala_{2}&=\tilde{A}^{\prime2}_{t}g_{xx}^{2}+\tilde{E}^{2}_{x}g_{uu}g_{xx}+(\tilde{B}_{x}^{2}+\tilde{B}^{2}_{z})
g_{tt}g_{rr}+(\tilde{A}^{\prime2}_{x}+\tilde{A}^{\prime2}_{y}+\tilde{A}^{\prime2}_{z})g_{tt}g_{xx}\,, \\
\cala_{4}&=\tilde{E}^{2}_{x}g_{xx}(\tilde{A}^{\prime2}_{y}+\tilde{A}^{\prime2}_{z})+\tilde{B}^{2}_{z}
(\tilde{A}^{\prime2}_{t}g_{xx}+\tilde{A}^{\prime2}_{z}g_{tt})
+\tilde{B}^{2}_{x}(\tilde{A}^{\prime2}_{x}g_{tt}+\tilde{E}^{\prime2}_{x}g_{uu}+\tilde{A}^{\prime2}_{t}g_{xx})\\
& +2\tilde{B}_{x}\tilde{B}_{z}\tilde{A}^{\prime}_{x}\tilde{A}^{\prime}_{z}g_{tt}
-2\tilde{B}_{z}\tilde{E}_{x}\tilde{A}^{\prime}_{t}\tilde{A}^{\prime}_{y}g_{xx}\,.
\end{split}
\end{equation}

By defining the conserved currents for the gauge fields $\tA_\m$ as
\begin{equation} \label{acons}
  \hJ^{\m}\equiv \frac{\partial\mathcal{L}}{\partial\tilde{A}^{\prime}_{\m}}\,,
\end{equation}
we have
\begin{equation}
\begin{split}
\hJ^{t}&=\frac{e^{-\phi}g_{xx}^{\frac{d-3}{2}}g_{\Omega\Omega}^{\frac{n}{2}}}{\sqrt{-{\rm det}\gamma}}
g_{xx}\left[-\tilde{B}_{z}\tilde{E}_{x}\tilde{A}^{\prime}_{y}+\tilde{A}^{\prime}_{t}(\tilde{B}^{2}_{x}
+\tilde{B}^{2}_{z}+g_{xx}^{2})\right]\,,\\
\hJ^{x}&=\frac{e^{-\phi}g_{xx}^{\frac{d-3}{2}}g_{\Omega\Omega}^{\frac{n}{2}}}{\sqrt{-{\rm det}\gamma}}
g_{tt}\left[\tilde{A}^{\prime}_{x}(\tilde{B}^{2}_{x}+g_{xx}^{2})+\tilde{B}_{x}\tilde{B}_{z}\tilde{A}^{\prime}_{z}\right] \,, \\
\hJ^{y}&=\frac{e^{-\phi}g_{xx}^{\frac{d-3}{2}}g_{\Omega\Omega}^{\frac{n}{2}}}{\sqrt{-{\rm det}\gamma}}
g_{xx}\left[\tilde{A}^{\prime}_{y}(\tilde{E}^{2}_{x}+g_{tt}g_{xx})-\tilde{A}^{\prime}_{t}\tilde{E}_{x}\tilde{B}_{z}\right] \,, \\
\hJ^{z}&=\frac{e^{-\phi}g_{xx}^{\frac{d-3}{2}}g_{\Omega\Omega}^{\frac{n}{2}}}{\sqrt{-{\rm det}\gamma}}
\left[g_{tt}\tilde{B}_{x}\tilde{B}_{z}\tilde{A}^{\prime}_{x}
+\tilde{A}^{\prime}_{z}(\tilde{B}^{2}_{z}g_{tt}+g_{xx}(\tilde{E}^{2}_{x}+g_{tt}g_{xx}))\right]\,,
\end{split}
\end{equation}

Then the on-shell action reads, in terms of currents,
\begin{equation} \label{ttt11}
S_{Dq}=-\mathcal{N}\int \dd^{d+2}\xi e^{-2\phi}g_{xx}^{d-2}g_{\Omega\Omega}^{n}\sqrt{-g_{tt}g_{rr}}
\frac{\zeta}{\sqrt{\zeta\chi-\frac{a_{1}^{2}}{g_{xx}^{2}+\tilde{B}^{2}_{x}}
-\frac{a_{2}^{2}}{g_{tt}g_{xx}+\tilde{E}^{2}_{x}}}} \,,
\end{equation}
where
\begin{equation} \label{ttt12}
\begin{split}
\zeta&=g_{tt}g_{xx}^{3}+\tilde{E}^{2}_{x}(\tilde{B}^{2}_{x}+g_{xx}^{2})
+g_{tt}g_{xx}(\tilde{B}^{2}_{x}+\tilde{B}^{2}_{z})\,,\\
\chi&=e^{-2\phi}g_{tt}g_{xx}^{d-1}g_{\Omega\Omega}^{n}
+{(\hJ^x)}^{2}+{(\hJ^y)}^{2}+\frac{g_{tt}g_{xx}}{g_{xx}^{2}+\tilde{B}^{2}_{x}}(\hJ^t)^{2}
+\frac{g_{tt}g_{xx}}{g_{tt}g_{xx}+\tilde{E}^{2}_{x}}{(\hJ^z)}^{2}\,, \\
a_{1}&=g_{tt}g_{xx}\tilde{B}_{z}{\hJ}^{t}-(g_{xx}^{2}+\tilde{B}^{2}_{x})
\tilde{E}_{x}{\hJ}^{y}\,,\\
a_{2}&=g_{tt}g_{xx}\tilde{B}_{z}{\hJ}^{z}+(g_{tt}g_{xx}+\tilde{E}^{2}_{x})\tilde{B}_{x}{\hJ}^{x}\,,
\end{split}
\end{equation}
The reality condition is
\begin{equation} \label{ttt13}
\chi(r_s)=a_{1}(r_s)=a_{2}(r_s) = 0 \,,
\end{equation}
where $r_s$ is the position such that (a singular shell condition)
\begin{equation}
  \zeta(r_s) = 0 \,.
\end{equation}
These four relations yield the conductivities
\begin{eqnarray}
\label{genc}
\begin{split}
\sigma^{xx}&=\frac{\tilde{B}^{2}_{x}+g^{2}_{xx}}{g_{xx}(\tilde{B}^{2}_{x}+\tilde{B}^{2}_{z}+g^{2}_{xx})}
\sqrt{e^{-2\phi}g_{xx}^{d-2}g_{\Omega\Omega}^{n}(g_{xx}^{2}+\tilde{B}^{2}_{x}
+\tilde{B}^{2}_{z})+({\hJ}^{t})^2}\,, \\
\sigma^{xy}&=\frac{\tilde{B}_{z}{\hJ}^{t}}{g_{xx}^{2}+\tilde{B}^{2}_{x}+\tilde{B}^{2}_{z}}\,, \\
\sigma^{xz}&=\frac{\tilde{B}_{x}\tilde{B}_{z}}{\tilde{B}^{2}_{x}+g^{2}_{xx}}\sigma_{xx}\,,
\end{split}
\end{eqnarray}
where $\s^{ij}$ is defined by $\hJ^i = \s^{ij} \tE_j$.
Note that the spacial dimensionality $d$ and internal space
enters only in $\s^{xx}$ and Hall conductivities are
independent of them.

\subsection{2+1 dimensioal anisotropic background}

In this subsection we focus on 2+1 field theory, which means $d=2$.
For an isotropic space, $g_{xx} = g_{yy}$, the result is simply obtained
by putting $B_x = 0$ in \eqref{genc}. However, we would like to
consider the possibility $g_{xx} \ne g_{yy}$ for
an anisotropic material or a model building for a strange metal.
For this purpose, we also need to introduce two electric fields $\tE_x$
and $\tE_y$ separately, since there is no rotational symmetry.

With the induced metric ($g_{xx} \ne g_{yy}$)
\begin{equation}
\dd s^{2}=g_{tt}\dd t^{2}+g_{rr}\dd r^{2}+g_{xx}\dd x^{2}+g_{yy}\dd y^{2}+g_{\Omega\Omega}\dd \Omega_{n}^{2}\,,
\end{equation}
and the gauge field
\begin{equation}
2\pi\a' A =\tilde{A}_{t}(r)\dd t + \left(-\tilde{E}_{x}t
+\tilde{A}_{x}(r)\right) \dd x
+\left(-\tilde{E}_{y}t+ \tilde{B}x+\tilde{A}_{y}(r)\right)\dd y \,,
\end{equation}
the DBI action \eqref{detgmn} reads
\begin{eqnarray}
S_{Dq}
&=&-\mathcal{N}\int \dd^{4}\xi e^{-\phi}g_{\Omega\Omega}^{\frac{n}{2}}\sqrt{-{\rm det}\gamma_{mn}}\,,
\end{eqnarray}
where
\begin{equation}
\begin{split}
-{\rm det}\gamma_{mn} &=g_{tt}g_{rr}g_{xx}g_{yy}+\tilde{E}_{x}^{2}g_{rr}g_{yy}++\tilde{A}^{\prime2}_{t}(\tilde{B}^{2}+g_{xx}g_{yy})
+\tilde{A}^{\prime2}_{x}g_{tt}g_{yy}+\tilde{A}^{\prime2}_{y}g_{tt}g_{xx}\\
& +\tilde{E}^{2}_{x}(g_{rr}g_{yy}+\tilde{A}^{\prime2}_{y})+\tilde{E}^{2}_{y}(g_{rr}g_{xx}+\tilde{A}^{\prime2}_{x})
+\tilde{B}^{2}g_{tt}g_{rr} \\
& -2\tilde{E}_{x}\tilde{E}_{y}\tilde{A}^{\prime}_{x}\tilde{A}^{\prime}_{y}
+2\tilde{A}^{\prime}_{t}\tilde{B}(\tilde{E}_{y}\tilde{A}^{\prime}_{x}-\tilde{E}_{x}\tilde{A}^{\prime}_{y}) \,.
\end{split}
\end{equation}
In terms of the conserved quantities defined by \eqref{acons}
\begin{equation}
\begin{split}
\hat{J}_{t}&=\frac{e^{-\phi}g_{\Omega\Omega}^{\frac{n}{2}}}{\sqrt{-\calg}}[\tilde{A}^{\prime}_{t}
(\tilde{B}^{2}+g_{xx}g_{yy})+\tilde{B}(\tilde{E}_{y}\tilde{A}^{\prime}_{x}-
\tilde{E}_{x}\tilde{A}^{\prime}_{y})]\,,\\
\hat{J}_{x}&=\frac{e^{-\phi}g_{\Omega\Omega}^{\frac{n}{2}}}{\sqrt{-\calg}}[
\tilde{A}^{\prime}_{x}(\tilde{E}^{2}_{y}+g_{tt}g_{yy})+\tilde{E}_{y}(\tilde{A}^{\prime}_{t}\tilde{B}
-\tilde{E}_{x}\tilde{A}^{\prime}_{y})]\,,\\
\hat{J}_{y}&=\frac{e^{-\phi}g_{\Omega\Omega}^{\frac{n}{2}}}{\sqrt{-\calg}}
[\tilde{A}^{\prime}_{y}(\tilde{E}^{2}_{x}+g_{tt}g_{xx})-\tilde{E}_{x}
(\tilde{A}^{\prime}_{t}\tilde{B}+\tilde{E}_{y}\tilde{A}^{\prime}_{x})]\,,
\end{split}
\end{equation}
the on-shell action reads
\begin{equation} \label{gen2p1}
S_{Dq}=-\mathcal{N}\int d^{4}\xi e^{-2\phi}g_{\Omega\Omega}^{n}\sqrt{-g_{tt}g_{rr}g_{xx}g_{yy}}
\frac{\zeta}{\sqrt{\zeta\chi-a^{2}}}\,,
\end{equation}
where
\begin{eqnarray}
\begin{split}
\zeta&=\tilde{B}^{2}g_{tt}+\tilde{E}^{2}_{x}g_{yy}+\tilde{E}^{2}_{y}g_{xx}+g_{tt}g_{xx}g_{yy}\,,\\
\chi&=e^{-2\phi}g_{\Omega\Omega}^{n}g_{tt}g_{xx}g_{yy}+\tilde{J}^{2}_{t}g_{tt}
+\tilde{J}^{2}_{x}g_{xx}+\tilde{J}^{2}_{y}g_{yy}\,,\\
a&=\tilde{E}_{x}\tilde{J}_{y}g_{yy}-\tilde{J}_{x}\tilde{E}_{y}g_{xx}-\tilde{B}\tilde{J}_{t}g_{tt}\,,
\end{split}
\end{eqnarray}
From a real-action condition
$$\zeta(r_s)=\chi(r_s)=a(r_s)=0 \,,$$
the conductivities are
\begin{eqnarray} \label{A1}
\begin{split}
\sigma^{xx}&=\frac{g_{yy}}{\tilde{B}^{2}+g_{xx}g_{yy}}\sqrt{
e^{-2\phi}g^{n}_{\Omega\Omega}(\tilde{B}^{2}+g_{xx}g_{yy})+\hJ^{2}_{t}}\,, \\
\sigma^{yy}&=\frac{g_{xx}}{\tilde{B}^{2}+g_{xx}g_{yy}}\sqrt{
e^{-2\phi}g^{n}_{\Omega\Omega}(\tilde{B}^{2}+g_{xx}g_{yy})+ \hJ^{2}_{t}}\,,\\
\sigma^{xy}&=-\sigma^{xy}=\frac{\tB\hJ_{t}}{\tB^{2}+g_{xx}g_{yy}}\,.
\end{split}
\end{eqnarray}

\subsection{The light-Cone AdS}

The geometry of the light-Cone AdS is explained
at the beginning of section \ref{LigthCone}.
Following~\cite{Kim:2010zq}, we introduce the worldvolume $U(1)$ gauge fields,
\begin{equation}
2\pi\alpha'A  = \left(\tE_{b}y + \th_{+}(r) \right) \dd x^+
+ \th_{-}(r) \dd x^-
+ \th_{y}(r) \dd y + \tB_b y \dd z\,,
\end{equation}
and field strengths
\begin{align}
&2\pi \alpha'F_{+y}=-\tE_{b} \,, \qquad 2\pi\a' F_{yz} = \tB_b \,, \\
& 2\pi \alpha' F_{+r}=-\th^{\prime}_{+}(r)\,, \quad
2\pi \alpha' F_{-r}=-\th^{\prime}_{-}(r)\,, \quad
2\pi \alpha'F_{ry}=\th^{\prime}_{y} \,.
\end{align}
The D7 brane DBI action is
\begin{eqnarray}
S_{D7} = -N_{f}T_{D7}\int \dd^{8}\xi\sqrt{-{\rm det}(g_{D7}+2\pi\alpha^{\prime}F)} = -\caln \int \dd^{5}\xi \, \mathcal{L} \ ,
\end{eqnarray}
where $\mathcal{N}\equiv2\pi^{2}N_{f}T_{D7}$.
In terms of conserved currentsdefined as
\begin{eqnarray}
\label{conc}
\hJ_{i}&=&\frac{\partial\mathcal{L}}{\partial \th^{\prime}_{i}} \,,
\end{eqnarray}
The Lagrangian reads
\begin{eqnarray}
 \call =  - g_{\Omega\Omega}^3 \sqrt{g_{--}g_{yy}g_{zz}g_{rr}^{D7}}
  \sqrt{\frac{ \mathfrak{g} \zeta^2}{
  \zeta\chi -g_{yy}g_{zz}\mathfrak{g} a_1^2
  -a_2^2 }} \,, \nonumber
\end{eqnarray}
where
\begin{equation}
\begin{split}
  g_{rr}^{D7}&=g_{rr}+R^{2}\theta^{\prime}(r)\,, \quad
    g_{\Omega\Omega}=R^{2}\cos^{2}\theta \\
  \zeta &= \left(-\mathfrak{g}(g_{yy}g_{zz} + \tB^2)+ \tE^{2}_{b}g_{zz}g_{--}\right) \,, \quad
  \mathfrak{g} = g_{+-}^2 - g_{++} g_{--} \,, \\
  a_1 &\equiv g_{--}\hJ_- + g_{+-} \hJ_{+} \,, \quad
  a_2 \equiv \tE_b g_{--}g_{zz} \hJ_z + \hJ_{+}\tB \mathfrak{g} \,, \\
  \chi &\equiv -\mathfrak{g}(g_{--}g_{yy}g_{zz}g_{\Omega\Omega}^3 + \hJ_+^2) + g_{--}g_{zz}\hJ_z^2 + g_{--}g_{yy}\hJ_y^2 \,.
\end{split}
\end{equation}
Note that there is always the point $r=r_s > r_H$ such that $\zeta = 0$ (a singular shell) where
\begin{eqnarray}
 \mathfrak{g}_s \equiv \left. \frac{\tE_b^2 g_{--}g_{zz}}{g_{yy}g_{zz}+\tB^2}\right|_{r=r_s} \,.
\end{eqnarray}
By requiring $a_1 = a_2 = \chi = 0$ at $r=r_s$ we have conductivities
\begin{equation} \label{RA_Bomsoo}
\begin{split}
  \hJ_- &= \left.\frac{g_{+-}}{g_{--}} \right|_{r=r_s} \hJ_+ \,, \\
  \hJ_z &=  \left. -\frac{\tB \hJ_+ \mathfrak{g}}{\tE_b g_{--}g_{zz}}
  = - \frac{\tB \hJ_+  }{g_{yy}g_{zz}+\tB^2}\right|_{r=r_s} \tE_b  \,, \\
  \hJ_y &= \left. \sqrt{\frac{g_{--} g_{zz} g_{\Omega\Omega}^3}{\tB^2 + g_{yy} g_{zz}}
  + \frac{g_{zz}^2 \hJ_+^2}{(\tB^2 + g_{yy}g_{zz})^2}} \right|_{r=r_s} \tE_b \,,
\end{split}
\end{equation}
where $ \hJ_{i} = \frac{J_{i}}{\mathcal{N} 2\pi \alpha'}$ and $ \tE_b = 2 \pi \alpha' E_b $.

\providecommand{\href}[2]{#2}

\begingroup

\raggedright

\endgroup

\end{document}